\documentclass[preprint]{aastex}

\shorttitle{Observational Tests}
\shortauthors{Young, Mamajek, Arnett \& Liebert}

\newcommand{\sol}{$M_\odot$}
\def \nuc#1#2{\relax\ifmmode{}^{#1}{\protect\text{#2}}\else${}^{#1}$#2\fi}

\begin{document}

\title{Observational Tests and Predictive Stellar Evolution}

\author{P. A. Young, E. E. Mamajek, David Arnett and James Liebert}
\affil{Steward Observatory, University of Arizona, 
933 N. Cherry Avenue, Tucson AZ 85721}
\email{payoung@as.arizona.edu, eem@as.arizona.edu, 
darnett@as.arizona.edu, jliebert@as.arizona.edu}

\begin{abstract}
We compare eighteen binary systems with precisely determined radii and masses 
from 23 to 1.1 \sol, and stellar evolution 
models produced with our newly revised code
TYCHO. ``Overshooting'' and rotational
mixing were suppressed in order to establish a baseline for isolating 
these and other hydrodynamic effects. 
Acceptable coeval fits are found for sixteen pairs
without optimizing for heavy element or helium abundance.
 The precision of these tests is limited by the accuracies 
of the observed effective temperatures.  
High dispersion spectra and detailed atmospheric modeling should give 
more accurate effective temperatures and heavy element abundances.
PV Cas, a peculiar early A system, EK Cep B, a known post-T 
Tauri star, and RS Cha, a member of a young OB 
association, are matched by pre-main sequence models. Predicted mass loss 
agrees with upper limits from IUE for CW Cep A and B. 
Relatively poor fits are obtained for binaries having at least one 
component in the mass range $1.7 < \rm M/M_\odot <2.6$, whose evolution is 
sensitive to mixing. These discrepancies are robust and consistent with
additional mixing in real stars. The predicted apsidal
 motion implies that massive star models are systematically less centrally
condensed than the real stars. If these effects are due to overshooting,
then the overshooting parameter $\alpha_{OV}$ increases with stellar mass.
The apsidal motion constants are controlled by radiative opacity 
under conditions close to those  directly measured in
laser experiments, making this test more stringent than possible before.
  
\end{abstract}


\section{INTRODUCTION}

Prior to any rigorous investigation of thermonuclear yields, mixing,
rotation, mass loss,
and other complex phenomena in the evolution of stars, it
is necessary to insure that the methodology used can reproduce
observations at the current state of the art.
Detached, double-lined eclipsing binaries provide the most accurate 
source of information on stellar masses and radii, and as such provide 
a crucial test for models of stellar evolution \citep{and91}. 

The predictive power of such simulations depends upon the extent to
which the validity of their oversimplifications can be tested. 
Phenomenology is particularly pernicious in that good tests, which
could give rise to falsification of the models and thus to progress,
can be nullified by parameter adjustment (``calibration'').
Similarly, the inclusion of new processes in the simulations is often
contentious in that there may be several candidates put forward as
the cause of the puzzling data. The treatment of the boundaries
of convection zones (``overshooting'') is ripe for re-examination
in terms of the underlying hydrodynamics \citep{asd99}. To aid this
development, we wish to understand just how well a standard convective 
treatment can do (i.e., we turn overshooting off). 
A more physically sound procedure will be presented subsequently.
Parameterized overshooting has been widely discussed (see 
\citet{mae75} for an early discussion, and \citet{spe97}
and \citet{andre,bfbc93,dcls99} 
for a recent ones with extensive references).
Rotation also may cause mixing \citep{mm00}.
Within the context of a plasma, rotation and convection may generate
magnetic fields, which by their buoyancy and angular momentum transport
may provide additional causes for mixing.

This paper represents a first step toward testing our extensively revised
stellar evolution and hydrodynamics code, TYCHO. Our goals are (1) the
understanding the predictive capability of stellar evolution theory by
the critical re-evaluation of its assumptions, and of its underlying 
basis in observations and in laboratory data,
(2) the examination of the tricky problem of mixing in stars to 
help design experiments
\citep{nova} and numerical tests \citep{asd99}, and 
(3) the development of an open source,
publicly available stellar evolution code with modern capabilities
for community use.

\subsection{Choice of Binaries}

The most comprehensive list of binary systems with accurately measured 
masses and radii is given in the review by \citet{and91}. 
A subset of the original binaries was chosen for this exploratory effort. 
The systems with the smallest uncertainties were picked such that the 
range of masses from 23 to 1.1 {\sol} was well sampled. The 
upper mass limit is the largest mass present in the data, while the 
lower limit is safely above the point at which the equation of state 
used in the modeling becomes inaccurate. This is primarily due to approximate
treatment of Coulomb contributions to the pressure; we use only the
weak screening limit for the plasma. The Coulomb 
interaction leads to a negative pressure correction of $\sim$ 8\%\ in 
the outer part of the convection region and $\sim$ 1\%\ in the core 
for a star of 1 {\sol} \citep{dap00}, and is less important for more
massive stars. We obtain a comparable
correction to the central pressure for a solar model (-1.7\%).

Aside from the exclusion of stars of still lower mass, 
no bias was applied in the selection process. 
For stars of $1 \rm\ M_\odot$ or less, the issue of possible
overshooting in the convective core is moot because their cores are
radiative. Also, because they rotate rapidly only for a brief part 
of their lives,
rotational mixing is expected to be less than for our selected stars.
The binaries used in the study along with their 
fundamental parameters are presented in Table~\ref{tbl-1}.
\citet{rib00} have revised the \citet{and91} temperature estimates,
and revised estimates for the masses of EM Car \citep{sti94} and 
CW Cep \citep{sti92} have been presented. \citet{lath96} have revised
the parameters for DM Vir.  All these changes are 
incorporated into Table~\ref{tbl-1}. 

\subsection{The Mixing Length}
If the purpose of a stellar evolution code is to make testable 
predictions of the behavior
of stars, then the adjustment of the mixing length by fitting stellar
data is repugnant. Alternatives are to constrain it by experiment or
by simulation. At present we know of no definitive experimental results
which determine the mixing length appropriate to stars,
although a variety of experiments do test other aspects of stellar
hydrodynamics and the codes used to simulate them \citep{nova}.
However, it is becoming possible to simulate turbulent, compressible
convection with sufficient realism to constrain the range allowed 
for the mixing length \citep{rosen,port94,port00,ell00,asi98}.
\citet{port00} find a mixing length $\alpha_{ML} = 2.68$ in units
of pressure scale height; this is based upon simulations having
mesh resolutions as high as $ 512 \times 512 \times 256 $ and
corresponding Rayleigh numbers as high as $ 3.3 \times 10^{12}$.
\citet{rosen} based their work on resolutions up to 
$ 253 \times 253 \times 163 $, but with a more realistic treatment of
radiative transfer and ionization. They were able to synthesize
the line profile of FeII $\lambda$5414 which compared well with
that observed. Their results agreed with standard 1-D models,
although they suggest that this might be ``the right result for the
wrong reason.'' Standard models use $\alpha_{ML} = 1.5$ to $2$,
which is significantly smaller than the
value of \citet{port00}.
For a red giant, \citet{asi98} inferred  $\alpha_{ML} = 1.6$, based
upon 2-D simulations but with fairly realistic microphysics.

\citet{can91,can92} have proposed a serious model to replace
mixing length theory; this has had the salutary effect of shifting
the debate to the physics of convection and away from the best choice
of mixing length. Finally, \citet{asd99} have shown from 2-D simulations
that the underlying physical picture for stellar convection is incomplete, 
even in the deep interior where the complication of radiative transfer
is minor. 

We simply choose $\alpha_{ML} = 1.6$, and look forward to the convergence
of these efforts to provide a convection
algorithm which is independent of stellar evolutionary calibrations.

\subsection{The TYCHO Code}
The evolutionary sequences
were produced with the TYCHO stellar evolution code. 
The code was originally developed for one dimensional (1D)
hydrodynamics of the late 
stages of stellar evolution and core collapse \citep{arn96}. It is
being completely rewritten as a general purpose, open source code
for stellar evolution and hydrodynamics. 
The present version is written in structured FORTRAN77 and is targeted
for Linux machines. It has been successfully ported to SunOS and SGI IRIX 
operating systems. It has extensive online graphics using PGPLOT, an open 
source package written by T. J. Pearson (\email{tjp@astro.caltech.edu}).
A library of analysis programs is being built (modules for
apsidal motion, pulsational
instability, reaction network links, and history of mass loss are now
available). The code is being put under source code control to 
allow versioning (this will allow particular versions of the code---for
example the one used in this paper, to be resurrected accurately at
later times), and to improve the reproducibility of results.

Knowledge of the radiative opacity of plasma at stellar conditions has
changed qualitatively in the past decade.  
Historically, solar and stellar atmospheres provided much of the empirical
data on hot plasmas. For example, \citet{kur} tabulates a range of effective
temperatures from 2,000 to 200,000~K. These temperatures and the corresponding
(low) densities characterized what was directly observable. Higher temperatures
could be found at lower densities (in non-LTE) or indirectly inferred by use 
of theoretical models. Terrestrial tests involved explosions which were
difficult to quantify with adequate precision to determine opacity.
Measurements of opacity in
a well characterized, hot, dense, laser-produced plasma have become
possible \citep{per91,spr92,mos95,dav00}.
The first experiments to simultaneously quantify temperature and
density with good precision \citep{per91,spr92} involved temperatures
$ T \approx 7 \times 10^5 \rm\ K$ and densities 
$\rho \approx 2 \times 10^{-2} \rm\ g\ cm^{-3}$, which 
are directly relevant to stellar evolution and to apsidal motion. 
The range of conditions which are experimentally accessible is expanding
with the development of new instruments and techniques. Not only can
direct measurements be made, but complex and sophisticated theories of
the physical state of the plasma can be tested, giving more reliable
extrapolations into conditions not yet experimentally accessible
\citep{per96,dav00}.
The conditions just quoted are encountered in stars of about $1 M_\odot$,
and are important for stellar evolution (see Ch.~7, \citet{arn96}).

The opacities used here are from \citet{opal96} and \citet{kur}, for a solar
abundance pattern \citep{ag88}. The 
 \citet{opal96} opacities were computed with 21 elemental species;
\citet{opal95} have shown that the remaining elements are so rare as to
have only a marginal
effect on the Rosseland mean opacities (for solar 
relative abundances of the heavier elements). While the OPAL opacities
were constructed for astrophysical use, the underlying experiments and
theoretical models are determined by the inertial confinement fusion
(ICF) community, reducing the danger of unconscious bias from astronomical
puzzles leaking back into the construction of opacities. Extension of the
opacity library to lower temperatures and lower entropies is planned.

TYCHO is designed to use an adaptable set of reaction networks; for these 
calculations, two networks were used. At higher temperatures
($T \ge 10^7 \rm\ K$), an 80 element reaction network was solved.
The reaction rates were from F. K. Thielemann (private communication);
see also \citep{fkt88}.
For lower temperatures this was replaced by a 15 element 
network which was designed for deuterium, lithium, beryllium, 
and boron depletion. The reaction rates were from \citet{cf88}.
The match at the temperature boundary was sufficiently good as
to require no smoothing. More recent compilations of nuclear
rates are available \citep{rau00,nacre}, but were not used here
to simplify the comparison with previous work.

The outer boundary condition was determined by use of 
the Eddington approximation to
a grey, plane parallel atmosphere, integrated in hydrostatic
equilibrium inward to a fitting point for the interior.
For the most extended model considered here, the ratio of mean
free path to radius was $\lambda/R \sim 5 \times 10^{-3}$, 
so that spherical
effects are negligible. The most vigorous mass loss considered
was so mild that the ram pressure $\rho v_{loss}^2$ at the
photosphere was $10^{-6}$ of the total thermal pressure, which is
consistent with the hydrostatic assumption. 
Such integrations were used to define the pressure and
temperature at the fitting points, $T_f(L,R)$ and  $P_f(L,R)$,
for the stellar luminosity $L$ and radius $R$. 
Their derivatives with respect
to stellar luminosity and radius were approximated by finite
differences constructed between three such integrations
(at $L,R$, $L+\delta L, R$, and $L,R+\delta R$).
Typically, of order 200 to 400 steps were used in the envelope
integration. 

We used Schwarzchild convection as formulated by \citet{kw90}, and
our treatment of convective overshooting was turned off.  

Mass loss was included and based on the theory of \cite{kud} for 
$T_{eff} \ge 7.5 \times 10^3 \ \rm K$ and the empirical approach
of \cite{dej} for lower effective temperatures. R. Kudritzki
kindly provided appropriate subroutines for the hotter regime.
Even for EM Car, the most massive system in the list, the effects 
of mass loss were modest ($0.6 \rm\ to\ 0.7\,  M_\odot$).

The equation of
state was that discussed in \citet{tim99}, augmented by the solution 
of the ionization equilibrium equations for H, He, and a set of heavier
elements scaled from the solar abundance pattern.
Both the equation of state and the thermonuclear reaction rates
are affected by coulomb properties of the plasma. Only weak screening
was necessary here.
Extension to include both weak and strong screening consistently 
in the equation of state and thermonuclear reaction rates is planned;
previous versions of the code included strong screening as well. 

Models were run for each mass, starting with a fully convective initial 
model on the Hayashi track and ending well beyond hydrogen depletion in the 
core. A more realistic approach would have been to form the stars by
accretion ( A. G. W. Cameron, private communication; \citet{nor00}).
We justify our choice by its simplicity, and by noting that only the last 
stages of the pre-main sequence are relevant here,
for which the two cases give similar results.

Zoning in the interior typically ranged from 200 to 500 zones.
All runs had solar heavy element abundance  \citep{ag88} and used a ratio of
mixing length to scale height of $\alpha =1.6$ for convection. 
This choice  gave a reasonably 
good solar model when compared to \citet{pison} and \citet{jcd}; 
inclusion of element settling by diffusion and
adjustment of the helium abundance would give improved consistency for
the present-day sun, but diffusion would have less time to operate in the 
more massive stars considered here. Rotational mixing was turned off.

\subsection{Related Investigations}
The stars are all at or near the main sequence, so that the possible
list of citations is enormous; the efforts of the Padua group 
\citep{bfbc93}, the Geneva group \citep{andre}, and the FRANEC
group \citep{dcls99} have comparable input physics and form a useful
context. 
 We focus discussion on \citet{polsb97}
and \citet{rib00}, who consider many of the same binaries,
and \citet{cg93} who examined the apsidal motion test.

The largest differences in microphysics between
\citet{polsb97} and this paper are our use of \citet{opal96}
rather than \citet{opal92}
opacities, a more realistic nuclear network, 
and a different approach to the equation of state, but none of
this seems to be particularly significant for the issues here. 
We do include mass loss, but these effects are not large.
\citet{polsb97} do not calculate the pre-main sequence evolution
(which is relevant to several of our binaries). 
They define an overshooting parameter which is fixed by previous
work on $\zeta$ Aurigae binaries \citep{spe97}.
They construct a grid of models in mass (0.5 to $40\rm\ M_\odot$)
and heavy element abundance ($ Z = 0.01$, 0.02 and 0.03), assuming 
$X = 0.76 - 3.0Z$ and $Y = 0.24 + 2.0 Z$ for the abundances of
hydrogen and helium. They minimize a $\chi^2$ error estimator
in four parameters: the masses $M_A$ and $M_B$, the age $t$ of
the binary, and $Z,$ the heavy element abundance. 

As a test of consistency for later evolution, we have reproduced
the $4$ and $8\rm\ M_\odot$ sequences of \citet{pols97}, which
did not use overshooting. The notoriously sensitive blue loops
were reproduced to graphical accuracy (their Figure 4) for the
same input physics. The codes seem highly consistent. 

\citet{rib00} used the models of \citet{cla95,cla97b} and 
\citet{cg95,cg98}, which use \citet{opal92} opacities and a 14 isotope network
and include overshooting and mass loss. 
They too interpolated in a grid, minimizing a  
$\chi^2$ error estimator. This procedure
was more complex than that used by \citet{polsb97}, and need
not be described here. Both heavy element abundance and helium abundance
were freely varied. 

Our strategy differs from both \citet{polsb97} and
\citet{rib00}, which may provide a useful contrast.
Here we are interested in isolating the possible inadequacy of
the standard formulation of stellar evolution, so we avoid
optimization of parameters as much as possible. 
By using (1) solar abundances and (2) the measured masses, 
we reduce the degrees
of freedom, and hopefully make the possible flaws in our
stellar evolution prescriptions easier to see.
By the same token, our models should fit the data less well
because we have not optimized abundances or masses.
Mathematically, optimization will almost always improve the fit,
but not necessarily for the correct reasons. However, 
the actual abundances may be different from our assumptions,
and the masses do have error bars.

\citet{cg93} used \citet{opal92} opacities, solar abundances,
a mixing length ratio $\alpha_{ML}=1.5$, and overshooting of
$\alpha_{OV}=0.2$ pressure scale heights (that is, essentially 
the same physics as the models used by \citet{rib00}), and 
computed structure constants for apsidal motion for seven of the
binaries we consider (EM Car, CW Cep, QX Car, U Oph, $\zeta$ Phe,
IQ Per, and PV Cas). 

Detailed comparisons will appear in the discussion below.

\section{FITTING MODELS TO BINARIES}

\begin{figure}
\figurenum{1}
\includegraphics[angle=-90,width=\textwidth]{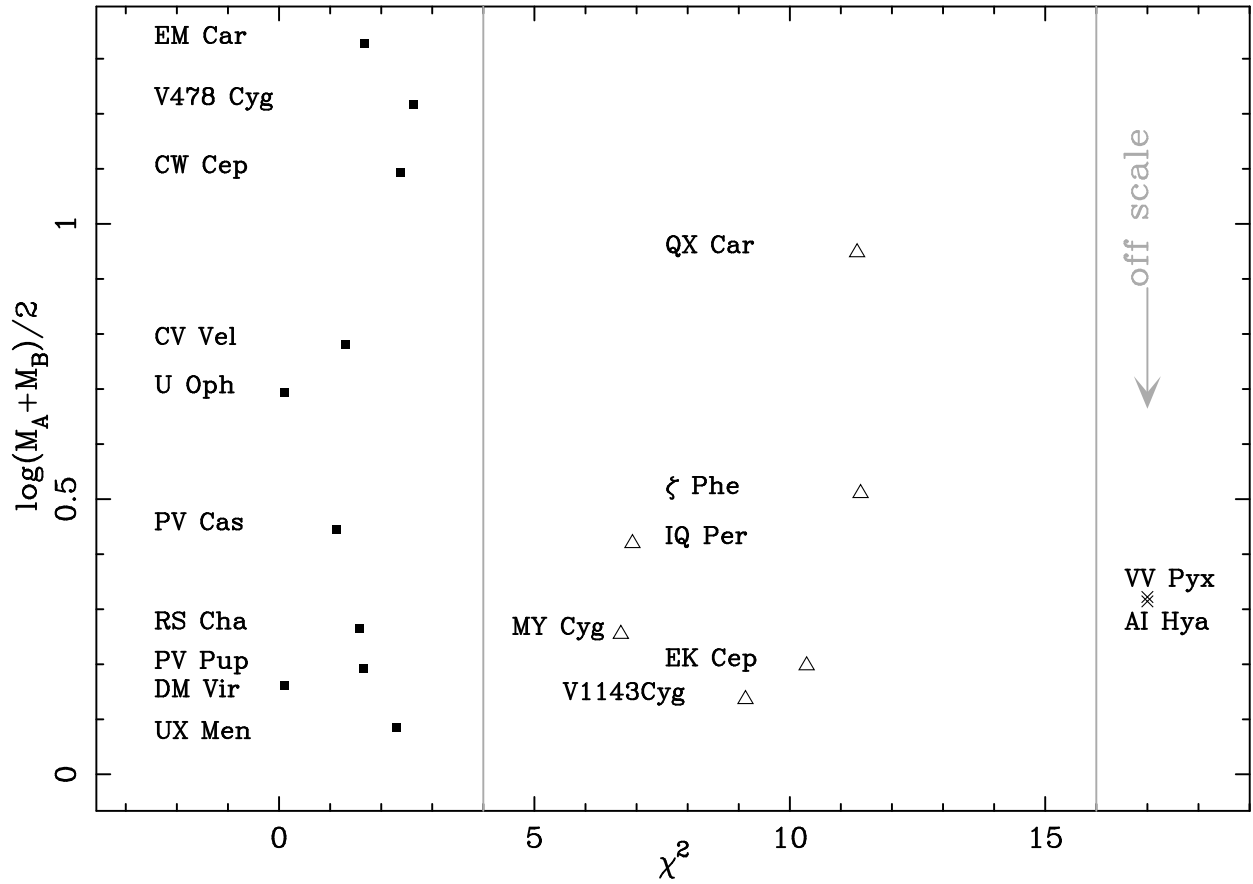}
\caption{$\chi^2$ for optimum models of selected binaries, versus mean mass
of the binary.}
\end{figure}
\placefigure{fig1}

The first step in comparing the binary data with the computations is
the choice of the best models. This was done by examining a
$\chi^2$ quantity for each binary pair, defined by
\begin{eqnarray}
\chi^2& = &((\log L(m_A,t)-\log L_{A})/\sigma_{L A})^2\nonumber \\
 &+& ((\log L(m_B,t)-\log L_{B})/\sigma_{L B})^2 \nonumber\\
&+&((\log R(m_A,t)-\log R_{A})/\sigma_{R A})^2 \nonumber\\
&+&((\log R(m_B,t)-\log R_{B})/\sigma_{R B})^2,
\end{eqnarray} 
where $A$ and $B$ denote the primary and the secondary
star, respectively. Here $L_A$ and $R_A$ are the observationally
determined luminosity and
radius of the primary, with $\sigma_{L A}$ and $\sigma_{R A}$
being the observational errors in $\log L_A$ and in $\log R_A$.
We convert the observational data for the radii to logarithmic form
for consistency. Correspondingly, $L(m_A,t)$ and $R(m_A,t)$
are the luminosity and radius of the model.
This $\chi^2$ was evaluated by computing two evolutionary
sequences, one for a star of mass $m_A$ and one for $m_B$, and storing
selected results from each time step. Then these files were marched
through, calculating $\chi^2$ at consistent times ($t_A=t_B$ to a 
fraction of a time step, which was a relative error of a few percent
at worst). The smallest
$\chi^2$ value determined which pair of models was optimum for that
binary.  Note that if the trajectories of both $A$ and $B$ graze 
the error boxes at the same time, $\chi^2\approx 4$. These error 
parameters along with the corresponding uncertainties from the 
observations are presented in Table~\ref{tbl-2}.

Figure 1 displays the
resulting  $\chi^2$ for each binary pair, in order of descending mean
mass. The binaries fall into three separate groups: ten have
excellent fits ( $\chi^2 < 4$; EM Car, V478 Cyg, CW Cep, CV Vel,
U Oph, PV Cas, RS Cha, PV Pup,  DM Vir and UX Men), 
six are marginal ($16 \leq \chi^2 \leq 4$;
QX Car,$\ \zeta$ Phe, IQ Per, MY Cyg, EK Cep, and V1143 Cyg),
and two are poor fits ( $\chi^2 > 16$, denoted {\em offscale} in
Figure 1;  VV Pyx and AI Hya).
The boundaries between these groups are indicated by vertical lines.

\subsection{Global Aspects of the Errors}

The weakness of a $\chi^2$ measure is that
it is most meaningful if the errors have a gaussian distribution
around the mean (\citet{numrec}, chapters 14 and 15), which does
not seem to be the case here. In particular, systematic shifts in
the empirical data, due to new analyses, can give significant shifts
in the error estimation.  \citet{rib00} have re-estimated the 
effective temperatures of 13 of the 18 binaries we have examined.
Five (QX Car, U Oph, PV Cas, AI Hya, and RS Cha) 
were changed by more than twice the error estimates of
either \citet{rib00} or \citet{and91}. 
Further, \citet{sti92} and \citet{sti94} have analyzed additional data
(from IUE) and find masses of CW Cep and EM Car which lie beyond twice
the error estimates. This is to be expected if
the errors are dominated by systematic effects, and warns us to distrust
all but our most robust inferences. 

Because the fractional errors in
mass and in radius are much more restrictive, it strongly supports the
need for renewed efforts to pin down the effective temperatures 
of these stars. The choice of $L$ and $R$ rather than $L$ and $T_{eff}$ 
in our definition of $\chi^2$ is significant: 
the smaller errors for $R$ make the $\chi^2$ more discriminating. 
\citet{polsb97} use $R$ and $T_{eff}$ which has the slight advantage
here of involving less propagation of observational errors, but
because $R$ is much more precise than $T_{eff}$, the effect is small
for the present data.

We have chosen to update the original data of \citet{and91}, incorporating
the changes made by the \citet{rib00} effective temperatures and the
\citet{sti92} and \citet{sti94} masses. We have used the new data for
DM Vir \citep{lath96}.
Our general conclusions are unaffected by which of these sets of
data we use.

\begin{figure}
\figurenum{2}
\includegraphics[angle=-90,width=\textwidth]{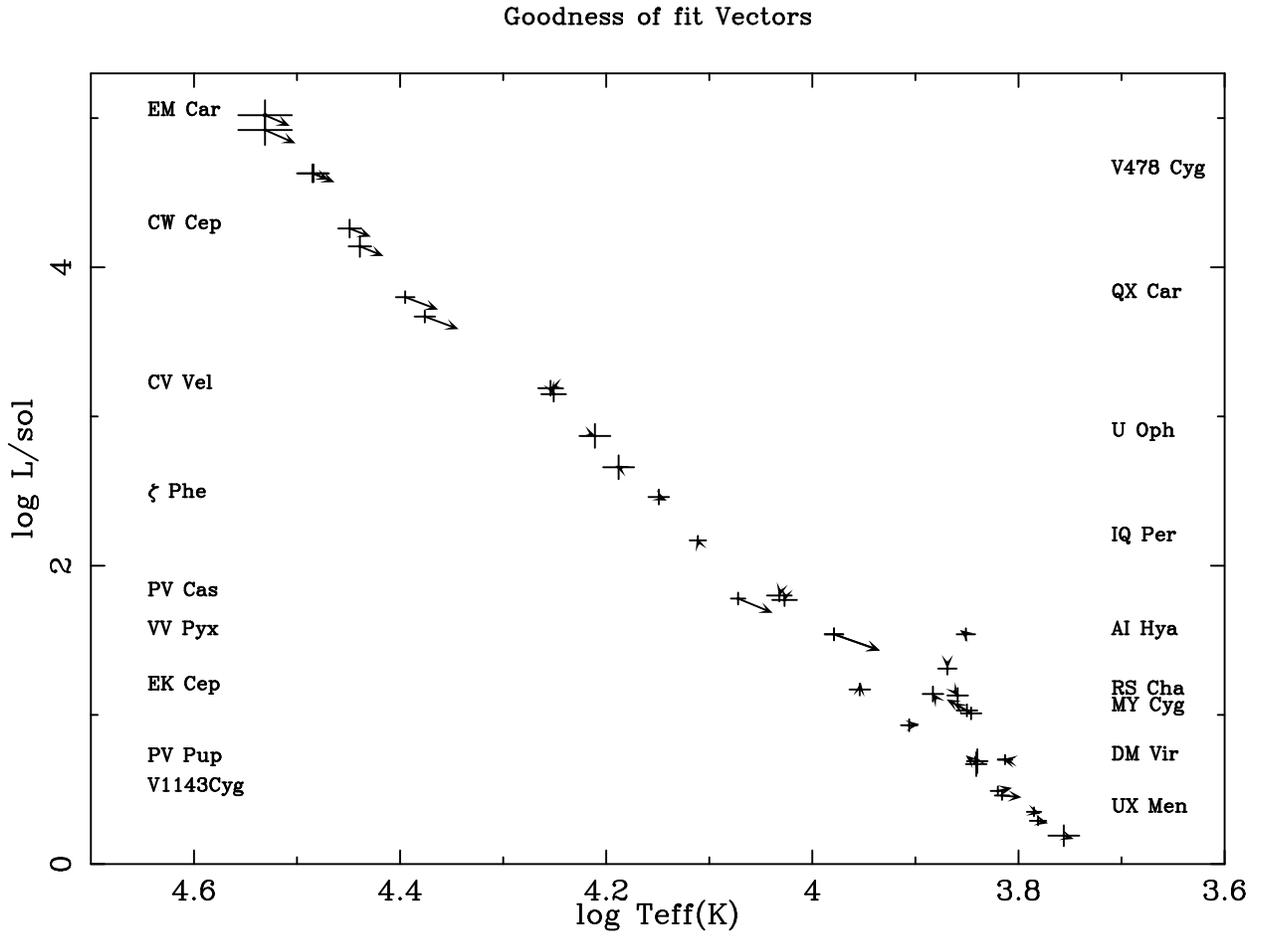}
\caption{Goodness of fit vectors for selected binaries, with observational
error bars.}
\end{figure}
\placefigure{fig2}

The comparison of observed and computed stars may be presented as an
goodness of fit vector, which has the advantage of 
being directly representable in the HR diagram for the stars. 
The observed points with error bars are plotted along with an arrow 
indicating the distance and direction to the best model point. 
The way in which the models differ from the observations 
can then be taken in at a glance. 
 Figure 2 shows the goodness of fit vectors, from the observed
points (shown with error bars) to the best model star (chosen as described
above). 
The largest discrepancy is the secondary of VV Pyx. Of the eight binaries
(QX Car, $\zeta$ Phe, IQ Per, VV Pyx, AI Hya, EK Cep, MY Cyg, V1143) 
which have mediocre or poor fits, seven (QX Car is the exception) have
at least one component lying in the range $1.7 < \rm M/M_\odot < 2.6$.
\citet{and90} noticed similar behavior.  

\begin{figure}
\figurenum{3}
\includegraphics[angle=-90,width=\textwidth]{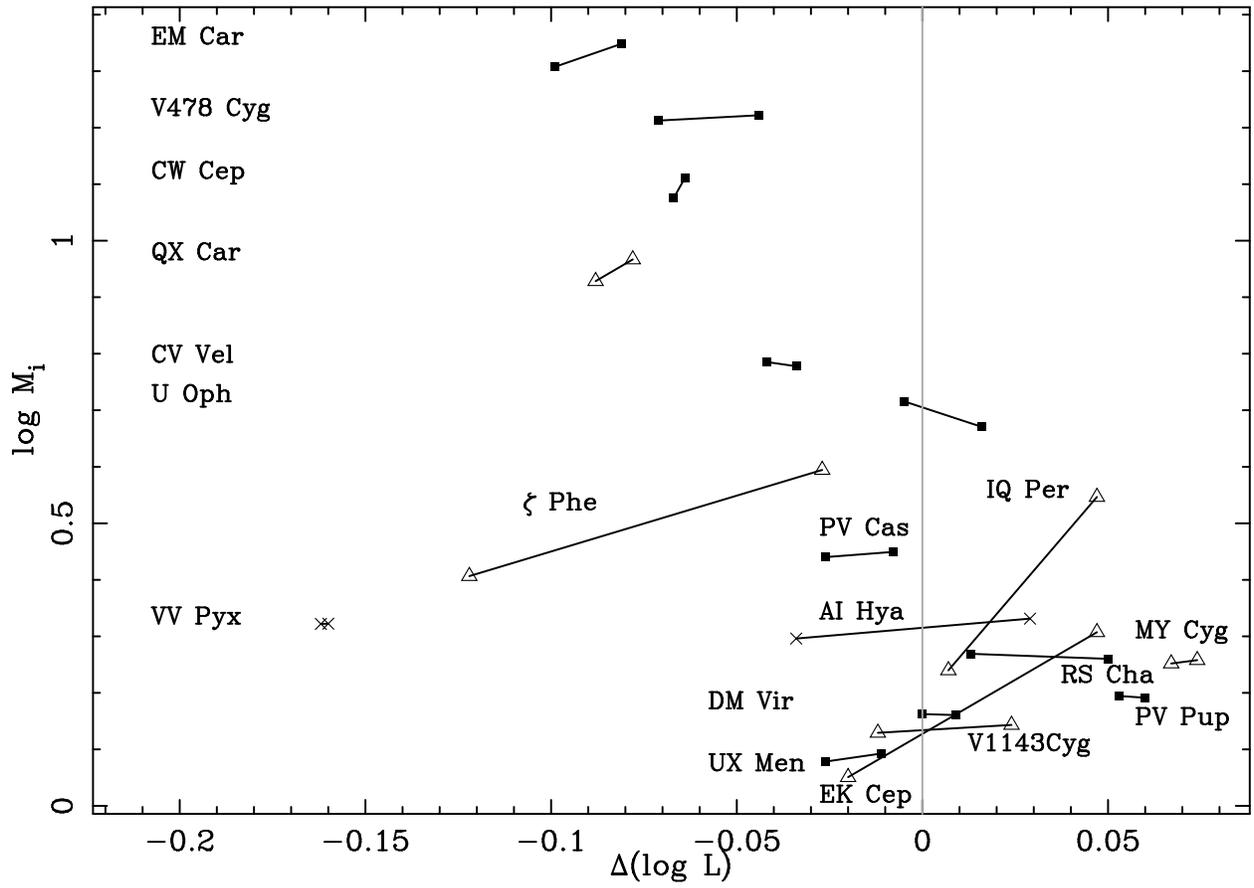}
\caption{Luminosity differences between best fit models and observations.}
\end{figure}
\placefigure{fig3}

Figure 3 shows the luminosity differences between the models
and the stars. The vertical axis is mass in solar units; binary components
are connected by a line.
The two binaries with $\chi^2 > 16$ (VV Pyx and AI Hya) 
are denoted by crosses; they are poor
fits and should be given little weight. Considering the best fits,
$\chi^2 < 4$ (solid squares), there is a dramatic trend:
the highest mass models (for example, EM Car) are
underluminous relative to the actual binaries, while the lower mass
models are not. 

Given the indications that the errors may be dominated by systematic
effects, we approach a statistical discussion with caution.
The two binaries which have $\chi^2 > 16$ are eliminated
from this statistical discussion 
on the basis that these fits are too poor to be meaningful.
In principle, the mean errors could show a systematic
shift in the models relative to the data, but because we choose
an optimum pair of models, the choice masks any absolute shift.
The error should reappear as a larger RMS difference.
For luminosity, the first moment of the difference between
model and stellar logarithmic luminosity is just the mean of this
difference, which  is $-0.017$ in the base ten logarithm
(the models are too dim by this amount).
The shift is smaller than the RMS error of the observational data,
which is 0.056. If there were a bad global mismatch, the RMS
difference in ``model minus star'' would be much larger than the average
error in the observations. However, the RMS difference between the
models and stars is 0.054, which is almost the same as the observational
error. The luminosity is basically a measure of the leakage time for
radiation, which is dominated by the value of the opacity in the radiative
regions. It samples the whole star, including the deep interior.
In a global sense, our mean leakage rate seems correct to the level of the 
statistical and observational error.

\begin{figure}
\figurenum{4}
\includegraphics[angle=-90,width=\textwidth]{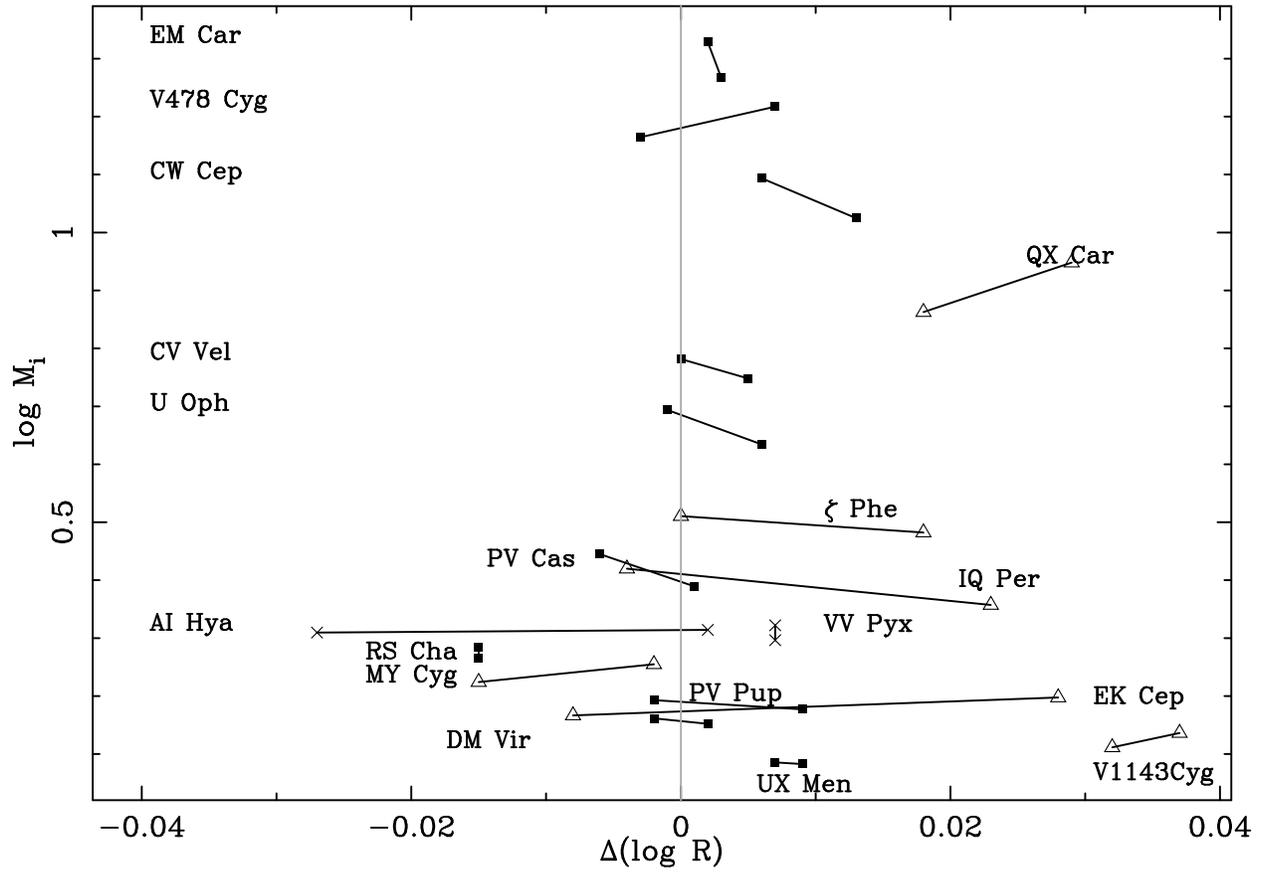}
\caption{Radius differences between best fit models and observations.}
\end{figure}
\placefigure{fig4}

The shifts in radius between the models and the stars are shown
in Figure 4. The vertical axis and the symbols are the
same as in the previous figure.
The mean shift is 0.0053 in the logarithm
(the models are too large by this small amount); the corresponding
standard error of the Andersen data for which the fits are
acceptable is 0.016, to be compared with an RMS differnce
between models and stars of 0.014.
Except for a few outlying cases, the distribution is fairly uniformly
distributed around zero. If only the best fits (squares) are considered, 
a subtle trend might be inferred: 9 of 12 of the  models 
above $4\rm M_\odot$ have radii which are too large. 

The corresponding mean shift in $\log T_{eff}$ is $-0.007$
(the models are too cool by this amount). Again, this is
small in comparison to the standard error of the observations ($0.014$).
The corresponding RMS difference between models and stars is $0.017$.
The effective temperature is a
surface quantity, and is more sensitive to the outer layers which
contain little of the stellar mass.

These numbers suggest that standard stellar evolutionary sequences 
of these stages should
be able to produce luminosities, radii, and effective temperatures within
11, 3 and 4 percent, respectively, of good observational
data. Otherwise, new physics is indicated. Because the standard
stellar evolutionary models do this well, small ``improvements'' may 
contain no information. We will emphasize systematic trends, and
those implications which emerge from several independent tests.

\subsection{Massive Binaries}

\begin{figure}
\figurenum{5}
\includegraphics[width=0.85\textwidth]{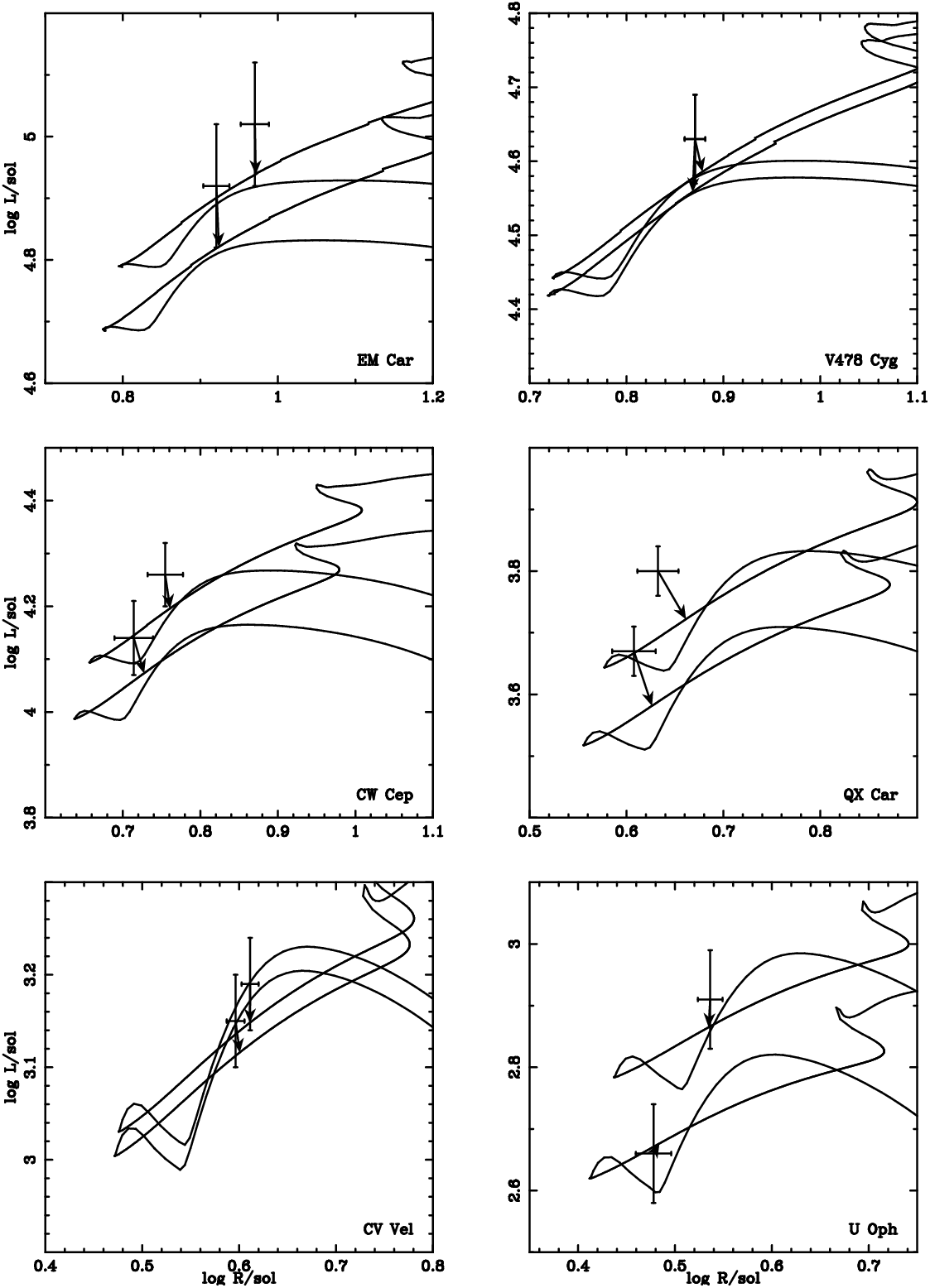}
\caption{Massive models: EM Car, V478 Cyg, CW Cep, QX Car, CV Vel and
U Oph. The masses range from 23 to 4.6 $\rm M_\odot$.}
\end{figure}
\placefigure{fig5}

Figure 5 shows the evolution of the model stars 
in log luminosity and log radius, for EM Car, V478 Cyg, CW Cep,
QX Car, CV Vel, and U Oph, corresponding to a mass  range 
from 23 to 4.6 $\rm M_\odot$. Except for QX Car ($\chi^2 = 11.3$),
models of these binaries have  $\chi^2 < 4$, and so represent good fits. 
The error bars are centered on the observed
stars; the arrows point from them to the optimum models. Notice
the the fits can be multivalued because the trajectories may pass
through the error boxes multiple times. This is shown occurring
first as the model descends from the pre-main sequence (pre-MS), and
again during main sequence hydrogen burning. If the stellar masses
are significantly different, this ambiguity is removed by the condition
that both components have the same age.
All the model stars are too dim (all the arrows point downward), a signal 
that the standard stellar evolution prescription is systematically wrong.

EM Car, being the most massive system, also has the most significant
mass loss. The model evolutionary sequences are set by the choice of initial 
mass, but the observational constraint on mass is applied after the best 
fitting model is determined. This is an implicit function of the choice of
initial mass, and iteration is required. Initial masses of 22.91 and
20.91 produce masses at fit of 22.25 and 20.12 $\rm M_\odot$,
respectively. This mass ratio of 0.904 is consistent with the
observational value of $0.910 \pm 0.011$ \citep{sti94}, and the masses
agree with observation to within the estimated errors 
($\pm 0.3\rm M_\odot$).

However, even this loss is still small.
A loss of $0.7 \rm\, M_\odot$ is about twice the uncertainty in mass 
determination, $ \pm 0.32\rm\, M_\odot$. Such a change in mass, 
since $L \propto M^4$, corresponds to a shift in luminosity
of $\Delta \log L \approx 0.05$, to be compared to the observational 
error in luminosity of $\Delta \log L = 0.1$, which is still larger.
This is due to the fact that effective temperature is less well
determined than the radius. A concentrated effort to refine the
effective temperature determinations for EM Car, V478 Cyg, CW Cep,
QX Car, CV Vel, and U Oph would translate directly
into much sharper constraints on massive star evolution. 
\citet{rib00} have revised the effective temperature for QX Car
(upward) by twice the quoted error, so that the inferred luminosity
increases. Prior to this revision, the fit to QX Car was good
($\chi^2 < 4$). This larger discrepancy for QX Car is in the same sense 
as noted for the other massive systems; the models are dimmer than the stars.
Taking the larger masses from \citet{and91}, with or without mass loss,
still gives good fits, and the models are still dimmer than the stars.
The result seems robust.

\citet{ppks} have used high resolution spectral images obtained
with the International Ultraviolet Explorer (IUE) to study the
winds from CW Cephei (HD 218066). They place upper limits on the 
mass-loss rates of
$1.0 \times 10^{-8}\rm \ M_\odot\rm\ yr^{-1}$ for the primary and
$0.32 \times 10^{-8}\rm \ M_\odot\rm\ yr^{-1}$ for the secondary. 
The model masses start at 12.9 and 11.9 and decrease only to 12.8 and
11.88 respectively, to be compared to $12.9 \pm 0.1$ and 
$ 11.9 \pm 0.1 \rm\ M_\odot$ \citep{sti92}.
The mass loss predicted by the \citet{kud} theory for our models
at the point of minimum error is
 $ 0.66\times 10^{-8}\rm \ M_\odot\rm\ yr^{-1}$ for the primary,
and $0.43 \times 10^{-8}\rm \ M_\odot\rm\ yr^{-1}$ for the secondary. 
If the upper limits were detections, this could be considered 
good agreement, considering the complexity of the 
problem of interpreting the system \citep{ppks}. The net loss of mass
up to this point is no larger than
the error in mass determination, $ \pm 0.1\rm\ M_\odot$.
Because the mass loss rate is restricted by these observations
to be at or below the
value we use, the mass loss process should have no larger effect
than we compute. Hence, the remaining discrepancy must come from
some other effect.

Table~\ref{tbl-3} gives the instantaneous mass loss rates from the
models, at the point of optimum fit, for the most massive binary
systems. At lower masses, the mass loss rates are smaller still.
Additional observational data on mass loss for these systems
could prove crucial in clarifying the role of mass loss in stellar 
evolution.
 
\citet{rib00} estimate ages for EM Car and CW Cep. 
Their procedure not only gives ages, but also error estimates
for those ages. Our ages agree with theirs to within these
errors, even though we use no overshooting and they do. It may be that the
convective region in high mass stars is sufficiently large that the gross
evolutionary properties of stars on the main sequence are not greatly affected
 by the overshooting correction. The understanding of the physics of
 overshooting is still too preliminary to do more than speculate on this issue.

\citet{polsb97} find acceptable fits for EM Car, V478 Cyg, CW Cep,
QX Car, and U Oph (it is probable that QX Car would not have
been a good fit with the revised effective temperatures),
but with increasingly lower heavy element abundances with
increasing mass. EM Car and V478 Cyg have fits at the limit of the
heavy element abundance range. \citet{rib00} find a similar effect: the
heavy element abundances of their massive binaries are marginally smaller
than those of the less massive ones.
Their effect is not quite as obvious as in \citet{polsb97}, 
perhaps because \citet{rib00} do not force the helium abundance 
to correlate with heavy element abundance, and it fluctuates for these systems.
The added degree of freedom may allow the fitting procedure to
obscure the trend.

This behavior could be interpreted as a galactic evolutionary effect, which
would be extremely interesting, but there is another possibility. 
The problem with the massive models
(see Figure 2 and  Figure 3) is that they are
too dim. Lower heavy element abundance gives higher luminosity because of 
reduced opacity. The fitting algorithms, having little freedom
for mass variation (thanks to the high quality of the data), 
must find lower heavy element abundance preferable, whether or not
the heavy element abundances are actually smaller. More effective mixing,
giving larger cores, also results in higher luminosities even if
the abundances are unchanged. {\em It is crucial to obtain spectroscopic
information to decide the issue.}
\citet{gui00} have recently examined V380 Cyg, which is a binary
of disparate masses
($\rm 11.1\pm 0.5\, M_\odot $ and $\rm 6.95\pm 0.25\, M_\odot $)
and evolutionary state. They conclude that more mixing is needed
($\alpha_{OV} = 0.6 \pm 0.1$). However, this system is complicated. Guinan 
et al. estimate that the system is approximately ten thousand years from
Roche lobe overflow by the primary. Thus, conclusions based on this system
should be made with caution.

The overshoot parameter $\alpha_{OV}$ may be a function of mass
(at least). Similar behavior can also be seen in 
\citet{cg91}, which finds different values of the best fit overshoot parameter
for five different masses. Alternatively, rotational mixing might be 
increasingly effective for larger masses.  Phenomenological prescriptions are 
valuable if they
capture the essential physics of the phenomena; if fitted parameters
turn out to be variable, a new formulation is needed.
 We have at least three causes for one effect;
sorting this out is an interesting theoretical and observational
challenge.

\subsection{Intermediate Mass Binaries}

\begin{figure}
\figurenum{6}
\includegraphics[width=0.85\textwidth]{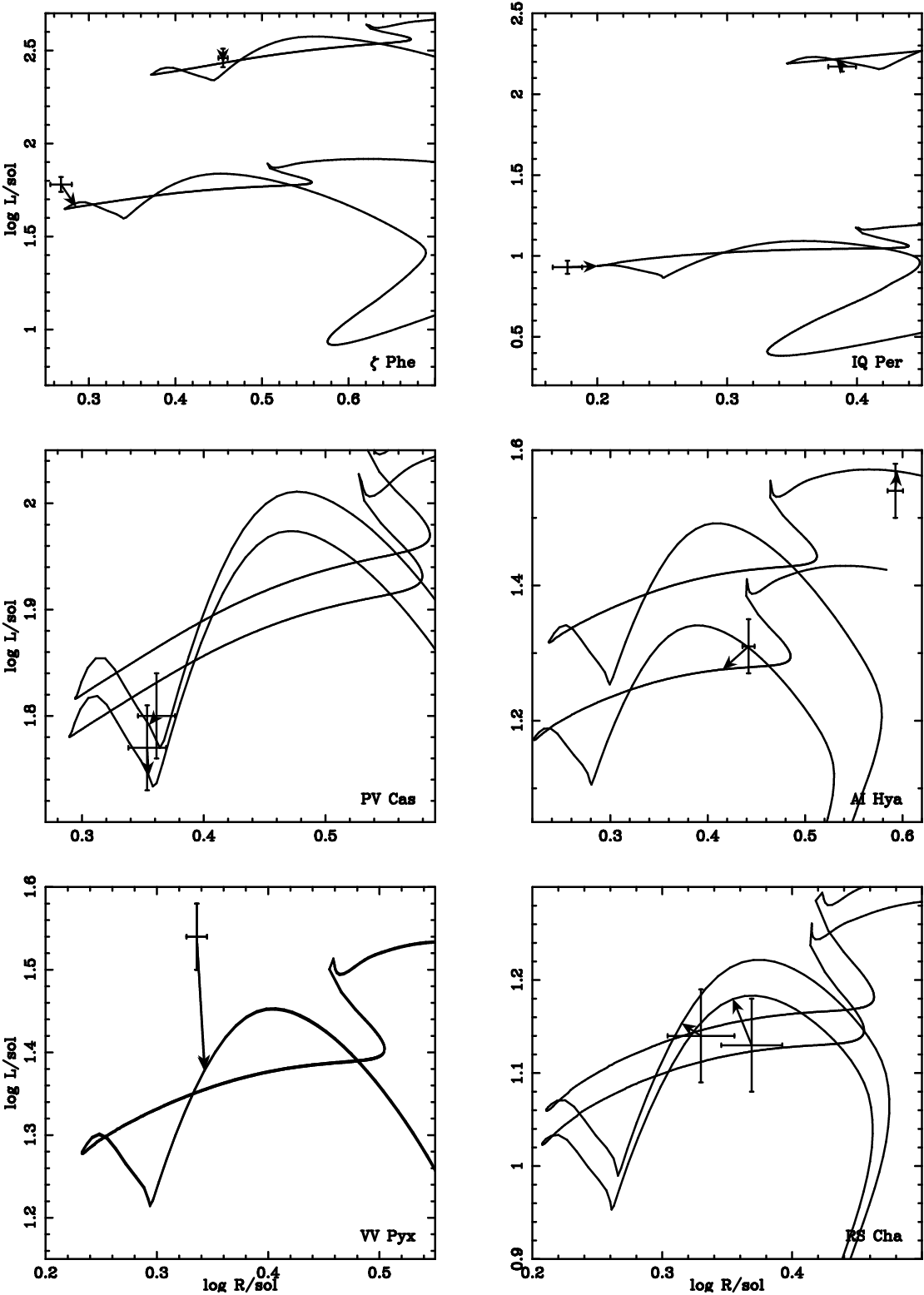}
\caption{Intermediate mass models: $\zeta$ Phe, IQ Per, 
PV Cas, AI Hya, VV Pyx and RS Cha. 
The masses range from 3.93 to 1.74 $\rm M_\odot$.}
\end{figure}
\placefigure{fig6}

Figure 6 shows $\zeta$ Phe, IQ Per, PV Cas, AI Hya, VV Pyx, and 
RS Cha, the group which has some of the most challenging binaries. 
The masses range from 3.9 to 1.1 $\rm M_\odot$.

Both $\zeta$ Phe and IQ Per have a mass ratio significantly different
from one: 0.65 and 0.49 respectively. Because the more massive components
will evolve more rapidly, common age is a stringent constraint.
In both cases, the error is dominated by the less massive component.
For  $\zeta$ Phe, the $2.55\rm\ M_\odot$ secondary is brighter than the
model; \citet{pols97} have the same problem. \citet{rib00} avoid it by
using a lower heavy element abundance (0.013) and a higher helium
abundance (0.29). The heavy element abundance might be tested
by high resolution spectroscopy. 
For IQ Per, the $1.74\rm\ M_\odot$ secondary
is too blue; its evolutionary track never gets so hot.
\citet{polsb97} attribute the difficulty in fitting $\zeta$ Phe and IQ Per
to problems in determining $T_{eff}$ of the secondary \citep{and91}, 
which is much dimmer because of the large mass ratios of the components. 

For $\zeta$ Phe and IQ Per, it is clear that much of our ``difficulty'' is
due to the relatively small error bars; see Figure 2. 
Consequently, small changes may improve the $\chi^2$ significantly,
even if they do not correspond to the physics of the system.
In the case of IQ Per, use of the \citet{rib00} value of effective 
temperature improves the fit, compared to \citet{pols97}, as does
adjustment of the abundances.

AI Hydrae is particularly interesting because the primary is fitted
by a model which is swiftly evolving, so that catching it in such a
stage is unlikely. Overshoot from the convective core would broaden 
the main sequence band and increase the age of the fast evolving primary, 
allowing  the possibility of a fit with a more probable, slower 
evolutionary stage. This is consistent with the conclusions of 
\citet{polsb97}, who 
find that AI Hya is the only binary for which the overshooting models 
give a greatly improved fit.
  
Some of the AI Hya behavior can be attributed to heavy element 
abundance effects. Both members are classified as peculiar metal line 
stars. The heavy element abundance of this system (from multi-color 
photometry) is 0.07 \citep{rib00}, which is 3.5 times the value used in 
the models. The true interior composition cannot be this metal rich. We have
examined a sequence which had a heavy element abundance of 0.03 rather
than 0.02. This modest change gave a dramatic shift toward lower 
luminosity ($\Delta \log L \approx 0.06$, which is three times the 
observational error) and cooler effective temperatures 
($\Delta \log T_e \approx 0.027$, which is also three times the 
observational error). However, if the heavy element abundance were
high only near the surface of the star, the opacity effects would 
produce a shift to the red in the evolutionary tracks, which 
would bring the models much more in line with the observations. 

VV Pyx has almost identical components, so that their coeval origin
has almost no effect on the fit. They track the same path at
essentially the same time. The fit is simply the point that the
observational error box is most closely approached, and should be
viewed with caution, especially as the models give a poor fit.

Only PV Cas  and RS Cha have good fits ($\chi^2<4$), and they are 
pre-main sequence (pre-MS). RS Cha has previously been suggested to be
in a pre-main sequence stage \citep{mam99}. The pre-MS identification 
would have important theoretical implications. If true, it implies 
that the error in the models occurs after the core convection has
been established in these stars. 
In any case, convection is an interesting possible cause for the problem; 
these binaries have at least one component with convective core burning.
PV Cas has sufficiently different masses to require us to examine the 
pre-MS fit seriously.

\subsection{Is PV Cas Pre-Main Sequence?}

Questions have been raised about the evolutionary status of 
PV Cas since \citet{pop87}. Previous attempts to fit the system to 
main sequence models \citep{pols97} have been unsatisfactory, mainly 
due to a large and irreconcilable age discrepancy between the members. 
Fitting both components to pre-MS models, however, produces excellent 
agreement. 

To test the case for PV Cas being pre-MS, we looked
for other observation clues. The double-lined eclipsing binary system
RS Cha was recently found by \cite{mam99,mam00} to be pre-MS. Not only were
pre-MS tracks for RS Cha a better fit than post-MS tracks, but 
two other observations strengthened the argument: (1) RS Cha
had several nearby {\it ROSAT} All-Sky Survey X-ray sources nearby
which were found to be very young, low-mass, weak-lined T Tauri 
stars, and (2) RS Cha's proper motion matched that of the T Tauri 
stars, suggesting a genetic tie. PV Cas is at a distance of 
660 pc \citep{pop87}, and a young stellar aggregate or membership within
an OB association could have been previously overlooked. 

Searching the Hipparcos and Tycho-2 catalogs, as well as examining 
PV Cas on the Digitized Sky Survey, we found no evidence for PV Cas being
a member of a known OB Association. More massive members of a putative
association would be included in the Hipparcos catalog with proper motions
similar to that given in the Tycho-2 entry for PV Cas, but none were found.
We searched for known groups of young stars with Vizier at CDS: 
the compilations of OB Associations by \citet{rup82}, \citet{me95}, and
\citet{dez99}, and open clusters by \cite{rup83} and \citet{lyn87}. 
The only possible known associations that PV Cas could belong to are
Cep OB3 (d=840 pc, $\Delta \theta =4.0\arcdeg,\ v_R = - 23\ \rm km/s$)
and Cep OB2 (d=615 pc, $\Delta \theta =9.5\arcdeg,\ v_R = - 21\ \rm km/s$),
however their projected separations from PV Cas are large (600 pc and
 100 pc, respectively), and their average radial velocities are far from 
Popper's value for PV Cas ($ v_R = - 3\ \rm km/s$). 
Hence, PV Cas does not appear to be connected to any known OB Associations
or clusters which help us to infer its nature.

The {\it ROSAT} All-Sky Survey (RASS) Bright Source Catalog (BSC; 
\citet{vog99}) 
and Faint Source Catalog (FSC; \citet{vog00}) were searched to
see whether there was any evidence for a clustering of X-ray-emitting
T Tauri stars in the vicinity of PV Cas. No concentration of sources
near PV Cas was detected, although the sensitivity of RASS at 660 pc
is about L$_x$ $\simeq$ 10$^{31}$ erg~s$^{-1}$, corresponding
to the very high end of the X-ray luminosity function for T Tauri stars
\citep{fei99}. Only one RASS-BSC source
was within 30\,\arcmin\ ($\sim$6 pc projected) 
of PV Cas, but its f$_{X}$/f$_{V}$ ratio was 2 magnitudes too high to be a 
plausible T Tauri star candidate. The only RASS-FSC source 
within 30\,\arcmin\ of PV Cas appeared to be 
related to a galaxy cluster on the Digitized Sky Survey. 

We conclude that we currently have no evidence for a pre-MS aggregate 
around PV Cas which could strengthen the argument for its pre-MS status.
However, the Taurus clouds also are forming low mass pre-MS stars without
high mass cluster counterparts.

\subsection{Lower Mass Binaries}

\begin{figure}
\figurenum{7}
\includegraphics[width=0.85\textwidth,keepaspectratio=true]{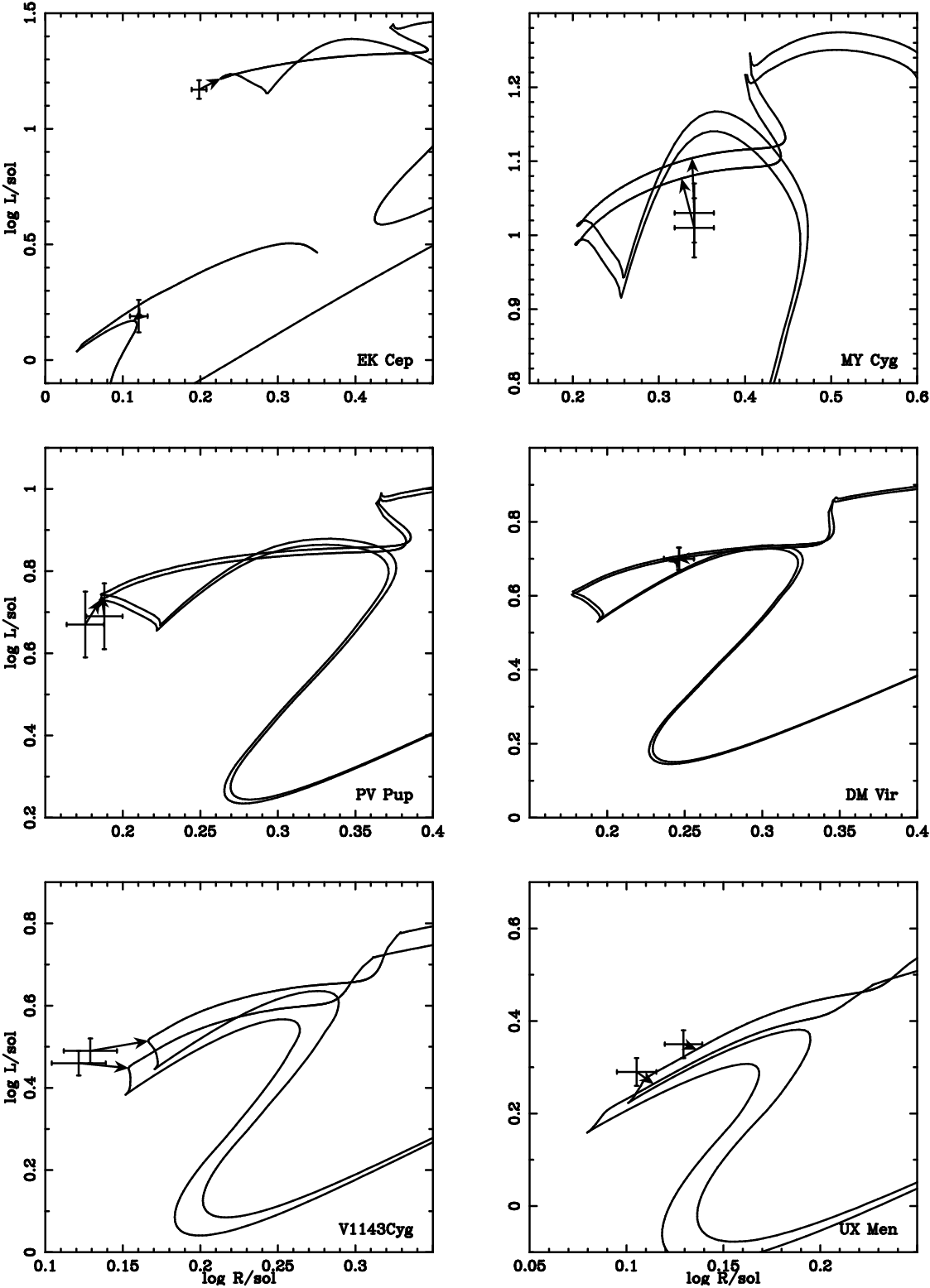}
\caption{Lower mass models: EK Cep, MY Cyg, PV Pup, DM Vir, 
V1143 Cyg, and UX Men. The masses range from 2.03 to 1.12 $\rm M_\odot$.}
\end{figure}
\placefigure{fig7}
Figure 7 shows EK Cep, MY Cyg, PV Pup, 
DM Vir, V1143 Cyg, and UX Men, whose
masses range from 2.03 to 1.12 $\rm M_\odot$. The \citet{rib00} effective
temperatures improve the fit for UX Men.

EK Cep has a large mass ratio. The optimum fit occurs as EK Cep B is 
still on the pre-MS track, in agreement with \citet{mar93} and \citet{cgm95}.
We find that the surface abundance of $\rm Li^6$ is depleted to
about $10^{-4}$ of its initial value, while  $\rm Li^7$ is depleted
from  $1.47 \times 10^{-9}$ to   $0.393 \times 10^{-9}$. This corresponds
to a depletion of elemental lithium of about 0.57 dex (base 10 logarithm).
This is somewhat larger than found by \citet{mar93} (0.1 dex), but may
be due to differences in the nuclear reaction rates used. 
In this range, the depletions are almost linear in the net cross section
for $\rm Li^7$ destruction. A careful analysis with a variety of rates 
is warranted:  \citet{mar93} suggest that the observations are in conflict 
with pre-MS models giving a Li depletion greater than 0.3 dex.

Although EK Cep has a large $\chi^2$ ($\chi^2 = 10.3$) 
if the radii are used in determining the fitting function, the situation 
is different for $\log L$-$\log T_{eff}$, the conventional HR plane. 
The observational errors are now larger, and the corresponding  
$\chi^2$ approaches 4. This confirms the importance of using the radii 
directly as a discriminant\citep{and91}.

MY Cyg and UX Men are found to be well into main sequence hydrogen
burning. MY Cyg is underluminous relative to the models. A higher
heavy element abundance would remove the discrepancy; observational
tests of this are needed. \citet{polsb97} found $Z=0.024$ and
\citet{rib00} found $Z=0.039$, which are consistent with this
suggestion.

PV Pup and V1143 Cyg are on the pre-MS/MS boundary. The fitting
procedure chooses the cusp at which the star settles down to main
sequence burning. This cusp shifts with small changes in abundance,
so these fits would benefit from independent measurement of the
abundances in these binaries.

DM Vir has been updated for \citet{lath96}. Although the changes were
small, the new fit is in the middle of main sequence hydrogen burning
instead of pre-MS contraction. The track lies well within the error bars;
the previous data also had $\chi^2 < 4$, although a much younger age
estimate.  Because the masses are almost equal, the coeval birth 
requirement has little effect, and the ages have a corresponding
uncertainty.

\subsection{Roche Lobes}

Observational selection favors binaries with a small separation. In
order to determine the true usefulness of these systems as tests of 
models of single star evolution it is necessary to know to what extent 
these systems are detached (noninteracting), and how far into the past 
and the future this condition is satisfied. 

In order to answer this question to first order for the systems in our 
sample, the average Roche lobe radius for each star was calculated using 
\begin{displaymath}
R_{Roche} = \frac{0.49a}{0.6+q^{-2/3}\ln(1+q^{1/3})}
\end{displaymath}
where $a$ is the binary separation and $q$ is the mass ratio with the
star in question in the numerator \citep{van95}. This average radius was 
then compared to the model radii to estimate when each star
overflows its Roche lobe. Dynamical evolution of the orbits was not taken 
into account. None of the models indicated significant mass transfer prior 
to the ages of the models closest to the observed points since early in the 
pre-Main Sequence evolution.

Four of the binaries in the sample have at least one member which, 
according to the model radii, will overflow their Roche lobes when 
they are between 1.3 and 2 times older than their current age. 
The results of the Roche lobe comparisons are given in Table~\ref{tbl-4}. All
stars labeled ``Post'' do not exceed their Roche Lobe radius until well 
into their post-main sequence evolution. Two stars are in contact early 
in the pre-main sequence evolution (EK Cep B and MY Cyg A). 
The times given for these stars 
correspond to when they contract below the critical radius and mass 
transfer ends. These numbers should be taken as a rough guideline at best, 
since the dynamical evolution of protostars 
is undoubtedly more complex than the simple algorithm used here. 
The models corresponding to the primaries EM Car and V478 Cyg 
exceed their Roche lobe radii in less than  $\rm{3\times 10^{6}}$\ years.

These values are approximate in that dynamical evolution is not
taken into account, the model radii do not match exactly the observed radii, 
and an approximate Roche lobe geometry was used to facilitate comparison 
to the spherically symmetric models.

\section{APSIDAL MOTION}

Apsidal motion in binaries allows us to infer constraints on the
internal mass distributions \citep{sch57}. Apsidal motion, that is,
rotation of the orientation of the orbital ellipse
relative to an inertial frame, does not occur
for binary orbits of point particles interacting by Newtonian gravity.
\citet{lc37} showed that the general relativistic expression for the
periastron shift of a double star is the same as for the perihelion shift 
of Mercury.  Following \citet{wei72} (see pages 194-7), the shift is
\begin{equation}
\big ( P/U \big )_{GR} = 3G(M_A+M_B)P/a (1-e^2)c^2,
\end{equation}
where $c$ is the speed of light and $G$ the gravitational constant.
Using units of solar masses and radii, and with the period $P$ in days, 
this dimensionless number becomes
\begin{equation}
\big ( P/U \big )_{GR} = 6.36 \times 10^{-6}(M_A+M_B)P/a (1-e^2),
\end{equation}
apsidal orbits per orbit. 
Tests of general relativity have reached high precision \citep{cwill};
the perihelion shift has now been tested to about $3 \times 10^{-3}$.
There has been some controversy as to a possible breakdown of general
relativity because of a discrepancy between observations and predictions
of the apsidal motion of some systems. This has been clarified by 
Claret (see \citet{cla97,cla98} for a recent discussion), who pointed out
errors in theoretical models and difficulties in observations, especially
for systems whose apsidal periods are too long for much to be measured
with modern equipment.  We adopt the point of view that general relativity
is better tested than subtleties in the evolution of binary stars, 
and ascribe errors to other causes (tidal effects not included,
rotational effects, and systematic errors in observational interpretation,
for example).

Tides induced by each companion give an additional interaction which
is not purely inverse square in the separation and cause apsidal motion.
\citet{qua96} have discussed the validity of the classical formula, which
we use, 
\begin{equation}
\big (P/U \big )_{CL} = (15/a^5)[ k_1 R_1^5 M_2/M_1 + k_2 R_2^5 M_1/M_2 ]f(e),
\end{equation}
where $P$ is the period of the orbit, $U$ the period of apsidal motion,
$M_i$ the mass and $R_i$ the radius of the star $i$, and
\begin{equation}
f(e) = (1 + {3 \over 2}e^2 + {1 \over 8}e^4)/(1-e^2)^5,
\end{equation}
where $e$ is the eccentricity of the orbit. The separation 
of the pair in solar radii is 
\begin{equation}
a = 4.207\ P^{2 \over 3} (M_1 + M_2)^{1 \over 3}, 
\end{equation}
if the period $P$ is
measured in days  and the masses in solar units. 
The classical apsidal motion formula gives accurate results
when the periods of the low-order quadrupole $g$, $f$ and $p-$modes are
smaller than the periastron passage time by a factor of about 10 or more
\citep{qua96}.
For EM Car, the lowest order pulsational mode of the primary has
 a period of $0.324$ days compared with the orbital period of 3.414 days and
an eccentricity of $0.0120\pm 5$, so that this condition is just satisfied.

If we assume that the observed apsidal motion is due only to these 
two effects, classical simple tides and general relativity, we have
\begin{equation}
\big ( P/U \big )_{OBS} - \big ( P/U \big )_{GR} = \big ( P/U \big )_{CL}.
\end{equation}
We use the products
$k_i R_i^5$ directly for greater precision, but quote the apsidal constants
$k_i$ for comparison.
\citet{pet95} has pointed out that accuracy problems may exist
because the relevant parameter is $ k_i R^5$, where $k_i$ is the 
apsidal constant and $R$ is the stellar radius, not just  $k_i$ alone.
 
\begin{figure}
\figurenum{8}
\includegraphics[angle=-90,width=\textwidth]{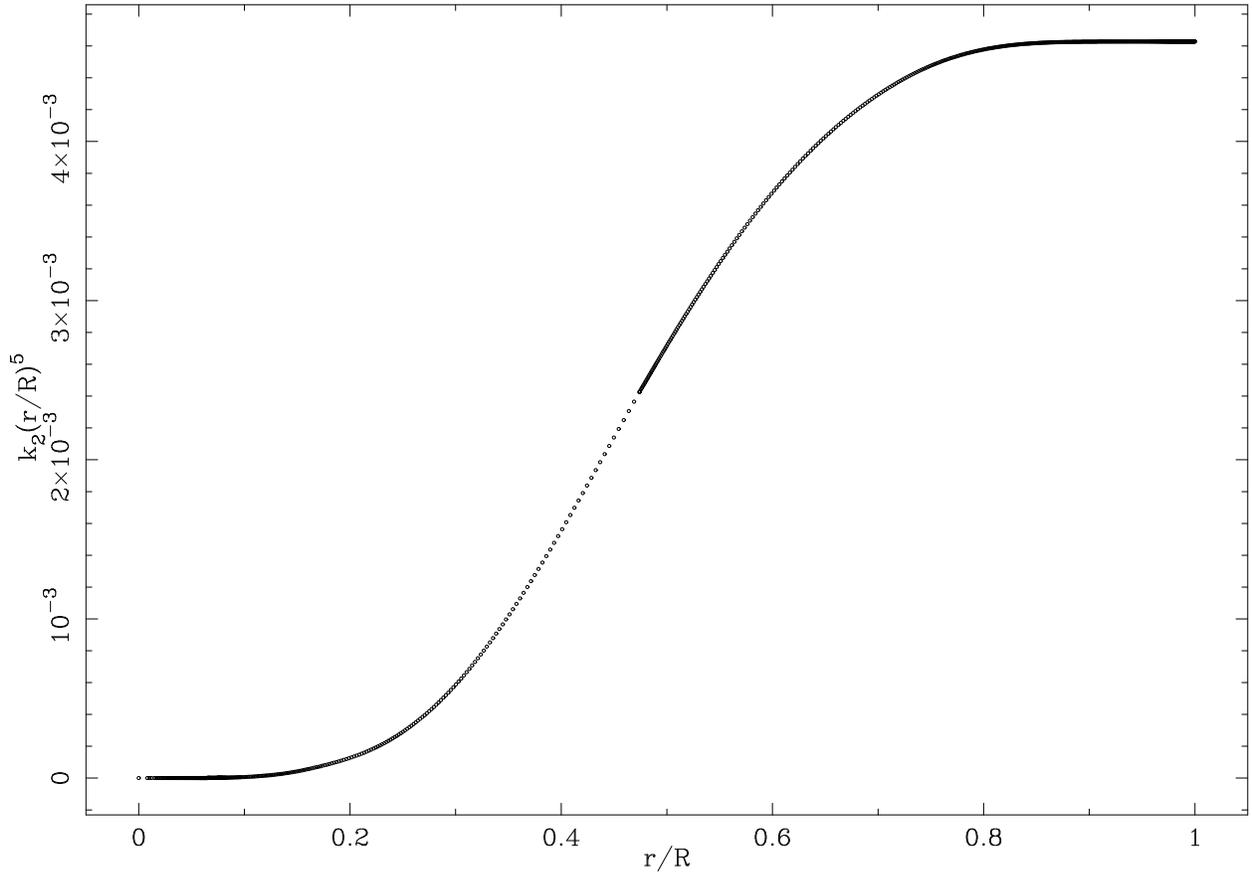}
\caption{Apsidal Constant integrand for EM Car primary. }
\end{figure}
\placefigure{fig8}

Figure 8 shows the integrand of the apsidal constant,
which approaches an asymptotic value as the integration exceeds 
about 0.7 of the radius. Inner regions contribute little because of their
small radii; outer regions have little mass.
The change from the interior (Henyey) integration to envelope integration
occurs around $r/R = 0.5$, and is visible in the change in the 
density of points. At the join, the temperature is about
$T \approx 6 \times 10^6\ \rm K$, 
and the density $\rho \approx 2.0 \times 10^{-2}\ \rm g cm^{-3}.$
This temperature is about ten times the value attained in the
early opacity experiments \citep{per91,per96} on the NOVA
laser, and is about half the goal for the National Ignition Facility (NIF).
For such main sequence (and pre-main sequence) stars, 
the apsidal constants are most
sensitive to the range of density and temperature which is directly
accessed by high energy density laser experiments 
(see \citet{nova} and discussion above). 
In this range, the new opacities show significant  deviation from those
previously used in astrophysics \citep{opal92,opal96}.

The \citet{pet99} catalog contains orbital elements for 128 binaries, 
including most (11 of 18) of the binaries in our list.   
Table~\ref{tbl-5} gives apsidal constants  $k_i$, as well
as the observed and predicted ratios of $P/U$. Given the significant
improvement in the opacities, a critical re-examination of these data
seems warranted.

\begin{figure}
\figurenum{9}
\includegraphics[angle=-90,width=\textwidth]{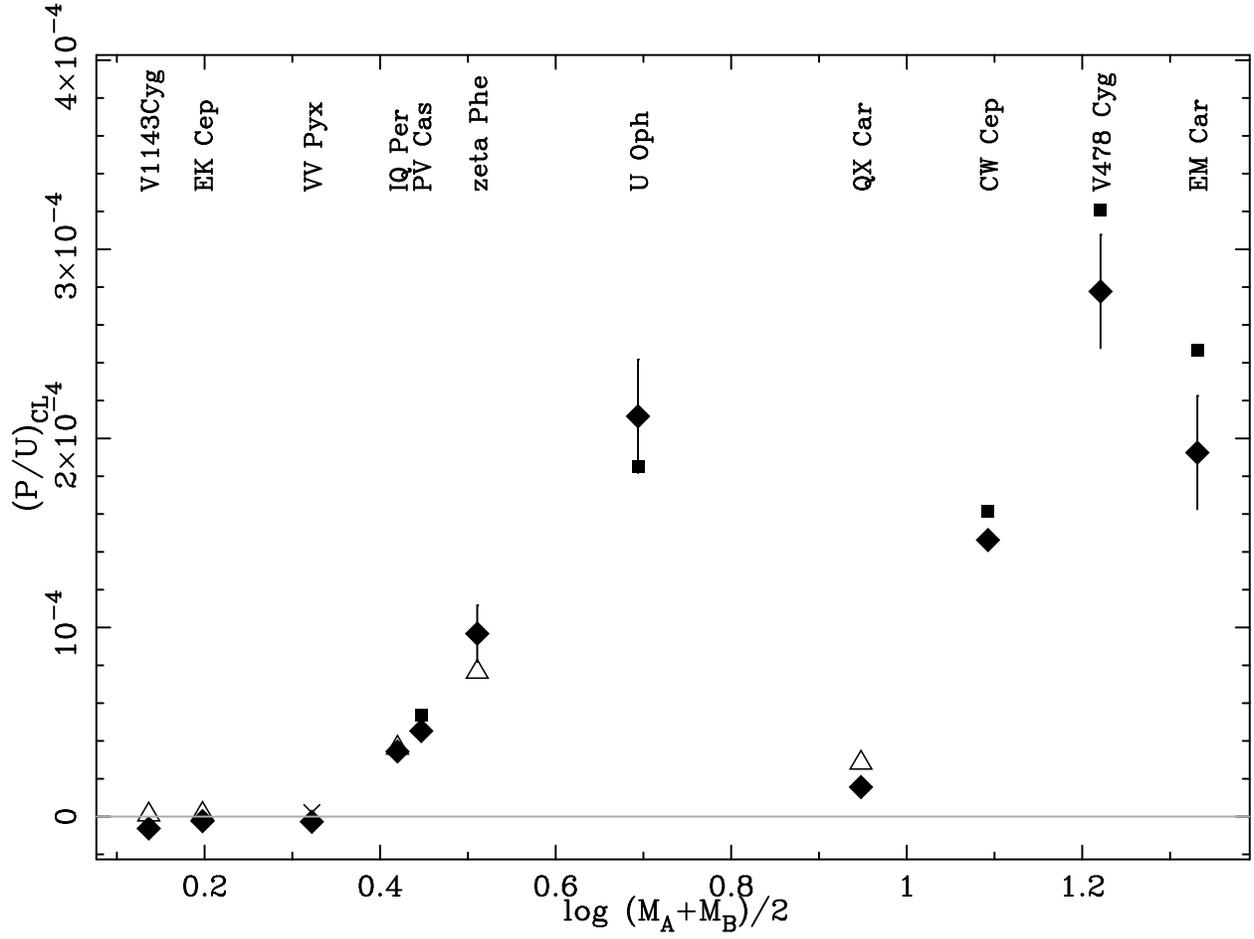}
\caption{Classical apsidal motion versus mean mass, for our binaries
with measured apsidal motion. $ (P/U)_{CL} = (P/U)_{OBS} - (P/U)_{GR} $
is assumed.}
\end{figure}
\placefigure{fig9}

Figure 9 shows the dimensionless rate of apsidal motion,
$ (P/U)_{CL} = (P/U)_{OBS} - (P/U)_{GR} $, which would be due to 
classical apsidal motion, plotted versus log of half the total binary 
mass. $P$ is the orbital period and $U$ the apsidal period.
The observational data (corrected for general relativity) are shown
as diamonds, with vertical error bars. The model predictions are
shown as solid squares ($\chi^2 < 4$) for the best fits, open triangles
for $4 < \chi^2 <16$, and crosses for $\chi^2 > 16$.
The massive binaries with good fits (EM Car, V478 Cyg, and CW Cep; 
$(M_A+M_B)/2 > 10 M_\odot$) have predicted apsidal motion
in excess of that observed, and QX Car also follows that trend.
These models are not as centrally condensed as the stars. 
This may be related to the underluminosity of these models found above.
Additional mixing would give more massive, convective cores, which
would result in both greater luminosity and more centrally condensed
structure.

Of the lower mass binaries with measured apsidal motion, only PV Cas
has a good fit model. Its predicted apsidal motion is also larger
than that observed (the stars are more centrally condensed). The other
binaries need better fitting models before the tests can be convincing.
Note that at the lowest tick mark in  Figure 9, the apsidal
period is measured in centuries.

\citet{cg93} have shown that inclusion of (a) overshooting, (b) variation in
heavy element abundance, and (c) rotation can produce models consistent with
the apsidal data; see also \citep{cla99}. This represents good progress
toward establishing the apsidal motion data as a useful test of 
stellar evolution. Our results, while not based upon identical assumptions,
are consistent. The challenge is that of correctly determining the
relative importance of the several different small effects which can 
give consistency with the observations. 

\section{CONCLUSIONS}

Standard stellar evolution, without embellishments such as overshooting
and rotation, does fairly well on these quantitative tests. While clear
discrepancies exist, they are relatively subtle. This makes it difficult
to uniquely identify exactly which additional physics is needed.
We find a detailed consistency with similar calculations by 
\citet{polsb97} and \citet{rib00}. It is important to test observationally
the abundance variations implied by $\chi^2$ optimization, as
such procedures may hide missing physics in parameter variation.
Laser experiments now explore the regions of temperature and density
which are relevant not only to conventional stellar evolution, but
also to apsidal motion tests.

Massive stars require more mixing than given by standard stellar
evolution, and probably more than the prescriptions for overshooting used 
by \citet{polsb97} and \citet{rib00}. Rotational mixing, overshooting
which is mass dependent, or something else is needed. Our mass loss
prescription is near the observational upper limit, so that additional
mass loss is an unlikely solution. 

Lower mass stars with convective cores are not well fit by standard
stellar evolution. Again, additional mixing is a promising answer.
Several of these binaries seem to be pre-main sequence; this will allow some
interesting tests of depletion of light nuclei and the mixing
processes.

We find two serious challenges: (1) disentangling conflicting solutions
of the relatively subtle discrepancies, and (2) controlling shifts in
the observational ``target areas'' due to systematic errors, which
seem to be larger than the statistical errors.
Improved determinations of effective temperature, and of heavy element
abundances (e.g., $\rm [Fe/H]$), would greatly improve these tests.

\acknowledgments
Stimulating discussions with Brigitta Nordstrom and Johannes Andersen 
are gratefully acknowledged. Special thanks are due to
 Marcel Arnould, Shimon Asida, John Bahcall, Audra Baleisis, Al Cameron,
Joergen Christensen-Dalgaard, Paul Drake, Rolf Kudritzki, Robert Kurucz,
T. J. Pearson, Ted Perry, Thomas Rauscher, Marc Rayet, Bruce Remington,
 Friedel Thielemann,  Frank Timmes, and Cliff Will,
whose work and/or advice contributed to aspects of this paper.
This work was supported in part by DOE, grant number DE-FG03-98DP00214/A001, 
and a subcontract from ASCI Flash Center at U. of Chicago.

\clearpage

\figcaption[gaus0.ps]{$\chi^2$ for selected binaries. \label{fig1}}

\figcaption[evector.ps]{Goodness of fit vectors for selected binaries.
 \label{fig2}}

\figcaption[gaus3.ps]{Luminosity differences. \label{fig3}}

\figcaption[gaus4.ps]{Radius differences.  \label{fig4}}

\figcaption[hr61.ps]{Massive models.  \label{fig5}}

\figcaption[hr62.ps]{Intermediate mass models.  \label{fig6}}

\figcaption[hr63.ps]{Lower mass models.  \label{fig7}}

\figcaption[apsidesema.ps]{Apsidal Constant integrand for EM Car primary.
  \label{fig8}}

\figcaption[puapside.ps]{Apsidal motion versus mass.
  \label{fig9}}

\clearpage
\begin{deluxetable}{crrllllll}
\tabletypesize{\scriptsize}
\tablecaption{Observed parameters for selected binary 
systems.\tablenotemark{a} \label{tbl-1}}
\tablewidth{0pt}
\tablehead{
\colhead{System} & \colhead{P(d)}   & \colhead{Star}   &
\colhead{Spect.} & \colhead{Mass/\sol} & \colhead{Radius/$R_\odot$} &
\colhead{$\log g{\rm (cm/s^2)}$}     & \colhead{$\log T_e{\rm(K)}$}  &
\colhead{$\log L/L_\odot$}           
}
\startdata
EM Car & $3.41$ & A & O8V  & $22.3\pm 0.3$\tablenotemark{b} &$9.34\pm 0.17$  &
 $3.864\pm 0.017 $\tablenotemark{b} & $4.531\pm 0.026$ & $5.02\pm 0.10 $  \\
HD97484 & \nodata & B & O8V  & $20.3\pm 0.3$\tablenotemark{b} &$8.33\pm 0.14$ &
 $3.905\pm 0.016 $\tablenotemark{b} & $4.531\pm 0.026$ & $4.92\pm 0.10 $   \\

V478 Cyg & $2.88$ & A & O9.5V  & $16.67\pm 0.45$ &$7.423\pm 0.079$  &
 $3.919\pm 0.015 $ & $4.484\pm 0.015$ & $4.63\pm 0.06 $  \\
HD193611 & \nodata & B & O9.5V  & $16.31\pm 0.35$ &$7.423\pm 0.079$  &
 $3.909\pm 0.013 $ & $4.485\pm 0.015$ & $4.63\pm 0.06 $   \\

CW Cep & $2.73$ & A & B0.5V &  $12.9\pm 0.1$\tablenotemark{c} & $5.685\pm 0.130$  &
 $4.039\pm 0.024$\tablenotemark{c} & $4.449\pm 0.011$\tablenotemark{d} & $4.26\pm 0.06$\tablenotemark{e} \\
HD218066 & \nodata & B & B0.5V &$11.9\pm 0.1$\tablenotemark{c}  & $5.177\pm 0.129$ &
$ 4.086\pm 0.024 $\tablenotemark{c} & $4.439\pm 0.011 $\tablenotemark{d} & $4.14\pm 0.07 $\tablenotemark{e} \\
\\
QX Car & $4.48$ & A & B2V &  $9.267\pm 0.122$ & $4.289\pm 0.091$  &
 $4.140\pm 0.020$ & $4.395\pm 0.009$\tablenotemark{d} & $3.80\pm 0.04$\tablenotemark{e} \\
HD86118 & \nodata & B & B2V &$8.480\pm 0.122$  & $4.051\pm 0.091$ &
$ 4.151\pm 0.021 $ & $4.376\pm 0.010 $\tablenotemark{d} & $3.67\pm 0.04 $\tablenotemark{e} \\

CV Vel & $6.89$ & A & B2.5V &  $6.100\pm 0.044$ & $4.087\pm 0.036$  &
 $4.000\pm 0.008$ & $4.254\pm 0.012$\tablenotemark{d} & $3.19\pm 0.05$ \\
HD77464 & \nodata & B & B2.5V &$5.996\pm 0.035$  & $3.948\pm 0.036$ &
$ 4.023\pm 0.008 $ & $4.251\pm 0.012 $\tablenotemark{d} & $3.15\pm 0.05 $ \\

U Oph & $1.68$ & A & B5V &  $5.198\pm 0.113$ & $3.438\pm 0.044$  &
 $4.081\pm 0.015$ & $4.211\pm 0.015$\tablenotemark{d} & $2.87\pm 0.08$\tablenotemark{e}\\
HD156247 & \nodata & B & B6V &$4.683\pm 0.090$  & $3.005\pm 0.055$ &
$ 4.153\pm 0.018 $ & $4.188\pm 0.015$\tablenotemark{d} & $2.66\pm 0.08 $\tablenotemark{e} \\
\\
$\zeta$ Phe & $1.67$ & A & B6V & $3.930\pm 0.045$ & $2.851\pm 0.015$ & 
$4.122\pm 0.009$ & $4.149\pm 0.010$\tablenotemark{d} & $2.46\pm 0.04$\tablenotemark{e} \\
HD6882 & \nodata & B & B8V & $2.551\pm 0.026$ & $1.853\pm 0.023$ & 
$4.309\pm 0.012$ & $4.072\pm 0.007$\tablenotemark{d} & $1.78\pm 0.04$\tablenotemark{e} \\

IQ Per & 1.74 & A & B8V & $3.521\pm 0.067$ & $2.446\pm 0.026$ & 
$4.208\pm 0.019$ & $4.111\pm 0.008$\tablenotemark{d} & $2.17\pm 0.03$\tablenotemark{e} \\
HD24909 & \nodata & B & A6V & $1.737\pm 0.031$ & $1.503\pm 0.017$ & 
$4.323\pm 0.013$ & $3.906\pm 0.008$\tablenotemark{d} & $0.93\pm 0.04$\tablenotemark{e} \\

PV Cas & 1.75 & A & B9.5V & $2.815\pm 0.050$\tablenotemark{d} & $2.297\pm 0.035$\tablenotemark{d} & 
$4.165\pm 0.016$\tablenotemark{d} & $4.032\pm 0.010$\tablenotemark{d} & $1.80\pm 0.04$\tablenotemark{e} \\
HD240208 & \nodata & B & B9.5V & $2.756\pm 0.054$\tablenotemark{d} & $2.257\pm 0.035$\tablenotemark{d} & 
$4.171\pm 0.016$\tablenotemark{d} & $4.027\pm 0.010$\tablenotemark{d} & $1.77\pm 0.04$\tablenotemark{e} \\
\\
AI Hya & 8.29 & A & F2m & $2.145\pm 0.038$ & $3.914\pm 0.031$ & 
$3.584\pm 0.011$ & $3.851\pm 0.009$\tablenotemark{d} & $1.54\pm 0.02$\tablenotemark{e} \\
+0\degr 2259 & \nodata & B & F0V & $1.978\pm 0.036$ & $2.766\pm 0.017$ & 
$3.850\pm 0.010$ & $3.869\pm 0.009$\tablenotemark{d} & $1.31\pm 0.02$\tablenotemark{e} \\

VV Pyx & 4.60 & A & A1V & $2.101\pm 0.022$ & $2.167\pm 0.020$ & 
$4.089\pm 0.009$ & $3.979\pm 0.009$\tablenotemark{d} & $1.54\pm 0.04$ \\
HD71581 & \nodata & B & A1V & $2.099\pm 0.019$ & $2.167\pm 0.020$ & 
$4.088\pm 0.009$ & $3.979\pm 0.009$\tablenotemark{d} & $1.54\pm 0.04$ \\

RS Cha & 1.67 & A & A8V   & $1.858\pm 0.016$ & $2.137\pm 0.055$ &
$4.047\pm 0.023$ & $3.883\pm 0.010$\tablenotemark{d} & $1.14\pm 0.05$\tablenotemark{e} \\
HD75747  & \nodata & B & A8V  & $1.821\pm 0.018$ & $2.338\pm 0.055$ &
$3.961\pm 0.021$ & $3.859\pm 0.010$\tablenotemark{d} & $1.13\pm 0.05$\tablenotemark{e} \\
\\
EK Cep & 4.43 & A & A1.5V & $2.029\pm 0.023$ & $1.579\pm 0.007$ & 
$4.349\pm 0.010$ & $3.954\pm 0.010$ & $1.17\pm 0.04$ \\
HD206821 & \nodata & B & G5Vp & $1.124\pm 0.012$ & $1.320\pm 0.015$ & 
$4.25\pm 0.010$ & $3.756\pm 0.015$ & $0.19\pm 0.07$ \\

MY Cyg & 4.01 & A & F0m & $1.811\pm 0.030$ & $2.193\pm 0.050$ & 
$4.007\pm 0.021$ & $3.850\pm 0.010$\tablenotemark{d} & $1.03\pm 0.04$\tablenotemark{e} \\
HD193637 & \nodata & B & F0m & $1.786\pm 0.025$ & $2.193\pm 0.050$ & 
$4.014\pm 0.021$ & $3.846\pm 0.010$\tablenotemark{d} & $1.02\pm 0.04$\tablenotemark{e} \\

PV Pup & 1.66 & A & A8V & $1.565\pm 0.011$ & $1.542\pm 0.018$ & 
$4.257\pm 0.010$ & $3.840\pm 0.019$ & $0.69\pm 0.08$ \\
HD62863 & \nodata & B & A8V & $1.554\pm 0.013$ & $1.499\pm 0.018$ & 
$4.278\pm 0.011$ & $3.841\pm 0.019$ & $0.67\pm 0.08$ \\
\\
DM Vir\tablenotemark{f} & 4.67 & A & F7V & $1.454\pm 0.008$ & $1.763\pm 0.017$ & 
$4.108\pm 0.009$ & $3.813\pm 0.007$ & $0.70\pm 0.03$ \\
HD123423\tablenotemark{f} & \nodata & B & F7V & $1.448\pm 0.008$ & $1.763\pm 0.017$ & 
$4.106\pm 0.009$ & $3.813\pm 0.020$ & $0.70\pm 0.03$ \\

V1143 Cyg & 7.64 & A & F5V & $1.391\pm 0.016$ & $1.346\pm 0.023$ & 
$4.323\pm 0.016$ & $3.820\pm 0.007$\tablenotemark{d} & $0.49\pm 0.03$\tablenotemark{e} \\
HD185912 & \nodata & B & F5V & $1.347\pm 0.013$ & $1.323\pm 0.023$ & 
$4.324\pm 0.016$ & $3.816\pm 0.007$\tablenotemark{d} & $0.46\pm 0.03$\tablenotemark{e} \\

UX Men & 4.18 & A & F8V & $1.238\pm 0.006$ & $1.347\pm 0.013$ & 
$4.272\pm 0.009$ & $3.785\pm 0.007$\tablenotemark{d} & $0.35\pm 0.03$\tablenotemark{e} \\
HD37513 & \nodata & B & F8V & $1.198\pm 0.007$ & $1.274\pm 0.013$ & 
$4.306\pm 0.009$ & $3.781\pm 0.007$\tablenotemark{d} & $0.29\pm 0.03$\tablenotemark{e} \\
\\

\enddata
\tablenotetext{a}{Detailed references and discussion may be 
found in \citep{and91}.}
\tablenotetext{b}{\citet{sti94}.}
\tablenotetext{c}{\citet{sti92}.}
\tablenotetext{d}{\citet{rib00}.}
\tablenotetext{e}{Adjusted here.}
\tablenotetext{f}{\citet{lath96}.}

\end{deluxetable}

\clearpage

\begin{deluxetable}{crrrrrrr}
\tabletypesize{\scriptsize}
\tablecaption{Parameters for selected binary systems. \label{tbl-2}}
\tablewidth{0pt}
\tablehead{
\colhead{System} & \colhead{Star}   & \colhead{Mass} & 
 \colhead{$\log R/R_\odot$} &
 \colhead{$\log T_e$}  &\colhead{$\log L$}     & \colhead{log Age (yr)} & 
 \colhead{$\chi^2$} 
}
\startdata
EM Car & A & $22.89$  & 0.972 & 4.509 & 4.933 & 6.666 & 1.67   \\
HD97484 & B & $21.43$  & 0.937 & 4.504 & 4.843 & 6.668 &         \\

V478 Cyg & A & $16.67$  & 0.879 & 4.464 & 4.566 & 6.807 & 2.67   \\
HD193611 & B & $16.31$  & 0.866 & 4.462 & 4.530 & 6.808 &         \\

CW Cep   & A & $13.52$   & 0.766 & 4.440 & 4.245 & 6.788 & 2.39  \\
HD218066 & B & $12.08$  & 0.720 & 4.421 & 4.075 & 6.798 &        \\
\\
QX Car   & A & 9.267     & 0.649 & 4.362 & 3.698 & 6.986 & 11.3  \\
HD86118  & B & 8.480     & 0.611 & 4.343 & 3.544 & 7.000 &        \\

CV Vel   & A & 6.100     & 0.614 & 4.231 & 3.103 & 7.604 & 1.30  \\
HD77464 & B & 5.996      & 0.603 & 4.228 & 3.070 & 7.607 &        \\

U Oph    & A & 5.198     & 0.538 & 4.198 & 2.820 & 7.687 & 0.43  \\
HD156247& B & 4.683      & 0.480 & 4.177 & 2.623 & 7.699 &        \\
\\
$\zeta$\ Phe& A & 3.930  & 0.457 & 4.136 & 2.409 & 7.831 & 11.4 \\
HD6882  & B & 2.551      & 0.283 & 4.028 & 1.633 & 7.836 &        \\

IQ Per   & A & 3.521     & 0.380 & 4.119 & 2.189 & 7.656 & 6.92 \\
HD24909 & B & 1.737      & 0.195 & 3.891 & 0.906 & 7.688 &        \\

PV Cas   & A & 2.827     & 0.362 & 4.015 & 1.736 & 6.576 & 1.12  \\
HD240208& B & 2.768      & 0.351 & 4.011 & 1.698 & 6.577 &        \\
\\
AI Hya   & A & 2.145     & 0.539 & 3.866 & 1.494 & 9.023 & 21.4\\
$+0\degr 2259$ & B &1.978& 0.395 & 3.865 & 1.204 & 9.025 &        \\

VV Pyx   & A & 2.101     & 0.349 & 3.920 & 1.331 & 6.850 & 32.2 \\
HD71581 & B & 2.099      & 0.350 & 3.920 & 1.330 & 6.850 &        \\

RS Cha   & A & 1.858     & 0.324 & 3.893 & 1.174 & 6.925 & 1.57  \\
HD75747 & B & 1.821      & 0.363 & 3.876 & 1.183 & 6.928 &        \\
\\
EK Cep   & A & 2.029     & 0.228 & 3.952 & 1.217 & 7.429 & 10.3 \\
HD206821& B & 1.124      & 0.086 & 3.761 & 0.166 & 7.432 &        \\

MY Cyg   & A & 1.811     & 0.347 & 3.851 & 1.052 & 9.117 & 6.69  \\
HD193637& B & 1.786      & 0.329 & 3.853 & 1.025 & 9.121 &        \\

PV Pup   & A & 1.565     & 0.183 & 3.858 & 0.750 & 7.980 & 1.66  \\
HD62863 & B & 1.554      & 0.182 & 3.855 & 0.738 & 8.101 &        \\
\\
DM Vir   & A & 1.460     & 0.241 & 3.802 & 0.639 & 7.177 & 0.15  \\
HD123423& B & 1.454     & 0.240 & 3.800 & 0.633 & 7.180 &        \\

V1143 Cyg& A & 1.391     & 0.169 & 3.789 & 0.447 & 7.323 & 9.13  \\
HD185912& B & 1.347      & 0.151 & 3.783 & 0.388 & 7.327 &        \\

UX Men   & A & 1.238     & 0.140 & 3.781 & 0.356 & 9.266 & 2.31  \\
HD37513 & B & 1.198      & 0.119 & 3.773 & 0.283 & 9.303 &        \\

\enddata 
\end{deluxetable}

\clearpage

\begin{deluxetable}{rcrr}
\tabletypesize{\scriptsize}
\tablecaption{Predicted instantaneous mass loss rates. \label{tbl-3}}
\tablewidth{0pt}
\tablehead{
\colhead{System} & \colhead{Star}   & \colhead{Mass} & 
\colhead{ Mass Loss Rate \tablenotemark{a}} 
}
\startdata
EM Car  	& A & $22.35$  & $1.82 \times 10^{-7} $ \\
        	& B & $20.51$  & $1.17 \times 10^{-7} $ \\
\\
V478 Cyg  	& A & $16.78$  & $2.50 \times 10^{-8} $ \\
        	& B & $16.47$  & $2.28 \times 10^{-8} $ \\
\\
CW Cep \tablenotemark{b} 	& A & $12.87$ & $0.66 \times 10^{-8}$ \\  
		& B & $11.88$ & $0.43 \times 10^{-8}$ \\
\\
QX Car 		& A & $9.257$ & $1.31 \times 10^{-9} $ \\  
		& B & $8.479$ & $6.32 \times 10^{-11} $ \\
\\
\enddata
\tablenotetext{a}{Predicted instantaneous mass loss rate in $M_\odot$/yr.}
\tablenotetext{b}{For IUE upper limit, see \citet{ppks}.}

\end{deluxetable}

\clearpage

\begin{deluxetable}{crrrrrrrr}
\tabletypesize{\scriptsize}
\tablecaption{Roche lobe parameters for selected binary systems. \label{tbl-4}}
\tablewidth{0pt}
\tablehead{
\colhead{System} & \colhead{Star}   & \colhead{Mass} & 
 \colhead{$a/ \rm R_\odot$} & \colhead{$e$} &
\colhead{$R/R_\odot$} & \colhead{$R_{roche}/R_\odot$} &
 \colhead{log Age (yr)} & \colhead{log Age(over)} 

}
\startdata
EM Car   & A & $22.89$ & 33.75 & $0.0120\pm5$ & 9.34 & 12.9  & 6.666 & 6.795 \\  
         & B & $21.43$ &       &              & 8.33 & 12.6 & 6.668 & 6.832 \\
V478 Cyg & A & $16.67$ & 27.31 & $0.019\pm2$  & 7.42 & 10.5 & 6.807 & 6.949 \\
         & B & $16.31$ &       &              & 7.42 & 10.2 & 6.808 & 6.965 \\
CW Cep   & A & $13.52$ & 24.22 & $0.0293\pm6$ & 5.68 & 9.33 & 6.788 & 7.085 \\
         & B & $12.08$ &       &              & 5.18 & 8.91 & 6.798 & 7.168 \\
QX Car   & A & $9.267$ & 29.82 & $0.0278\pm3$ & 4.29 & 11.5 & 6.986 & Post  \\
         & B & 8.480   &       &              & 4.05 & 11.0 & 7.000 & Post  \\
CV Vel   & A & 6.100   & 34.97 & $<4\times 10^{-3}$ & 4.09 & 17.0 & 7.604&Post\\
         & B & 5.996   &       &              & 3.95 & 13.2 & 7.607 & Post  \\
U Oph    & A & 5.198   & 12.76 & $0.0031\pm2$ & 3.44 & 4.90 & 7.687 & 7.910 \\
         & B & 4.683   &       &              & 3.01 & 4.68 & 7.689 & 8.030 \\
$\zeta$\ Phe& A & 3.930 & 11.04 & $0.0113\pm20$ & 2.85 & 4.57 & 7.831 & 8.276 \\
         & B & 2.551   &       &              & 1.85 & 3.80 & 7.836 & 8.797  \\
IQ Per   & A & 3.521   & 10.58 & $0.076\pm4$  & 2.45 & 4.68 & 7.656 & 8.405 \\
         & B & 1.737   &       &              & 1.50 & 3.39 & 7.688 & Post  \\
PV Cas   & A & 2.827   & 10.85 & $0.032\pm1$  & 2.30 & 4.07 & 6.576 & 8.669 \\
         & B & 2.768   &       &              & 2.26 & 4.07 & 6.577 & 8.697  \\

AI Hya   & A & 2.145   & 27.63 & $0.230\pm2$  & 3.91 & 10.7 & 9.023 & Post  \\
         & B & 1.978   &       &              & 2.77 & 10.2 & 9.025 & Post  \\
VV Pyx   & A & 2.101   & 18.77 & $0.0956\pm9$ & 2.17 & 7.08 & 6.850 & 9.066 \\
         & B & 2.099   &       &              & 2.17 & 7.08 & 6.850 & Post  \\

RS Cha   & A & 1.858   & 9.14  & $0.030\pm15$ & 2.14 & 3.47 & 6.925 & Post  \\
         & B & 1.821   &       &              & 2.34 & 3.47 & 6.928 & Post  \\
EK Cep   & A & 2.029   & 16.64 & $0.190\pm3$  & 1.58 & 7.08 & 7.429 & Post  \\
         & B & 1.124   &       &              & 1.32 & 5.50 & 7.432 & 5.669 \\

MY Cyg   & A & 1.811   & 16.27 &       & 2.19 & 6.17 & 9.117 & 5.646\\
         & B & 1.786   &       &              & 2.19 & 6.17 & 9.121 & Post \\
PV Pup   & A & 1.565   & 8.62  & $0.050\pm1$  & 1.54 & 6.17 & 7.980 & Post \\
         & B & 1.554   &       &              & 1.50 & 6.17 & 8.101 & Post \\

DM Vir   & A & 1.460   & 16.79 & $< 10^{-4}$  & 1.73 & 6.31 & 7.177 & Post \\
         & B & 1.454   &       &              & 1.73 & 6.31 & 7.180 & Post \\
V1143 Cyg& A & 1.391   & 22.83 & $0.540\pm5$  & 1.35 & 8.71 & 7.323 & Post \\
         & B & 1.347   &       &              & 1.32 & 8.51 & 7.327 & Post \\
UX Men   & A & 1.238   & 14.69 & $0.015\pm17$ & 1.35 & 5.62 & 9.266 & Post \\
         & B & 1.198   &       &              & 1.27 & 5.50 & 9.303 & Post \\
\enddata
\end{deluxetable}

\clearpage

\begin{deluxetable}{rcrrrrrrr}
\tabletypesize{\scriptsize}
\tablecaption{Apsidal comparisons for selected binary systems. \label{tbl-5}}
\tablewidth{0pt}
\tablehead{
\colhead{System} & \colhead{Star}   & \colhead{Mass} & \colhead{$-\log k_i$} & 
  \colhead{$(k_2R^5)$ \tablenotemark{a}} &
\colhead{$\rm P/U_{CL}$ \tablenotemark{b}}    & 
\colhead{$\rm P/U_{GR}$ \tablenotemark{b}}    &
\colhead{$\rm P/U_{CL+GR}$ \tablenotemark{b}} &
\colhead{$\rm P/U_{OBS}$ \tablenotemark{b}} 
}
\startdata
EM Car  & A & $22.35$  & 2.240 & $437.9$ & $2.46 $ & $0.275$ &
 2.74 & $2.2 \pm 0.3$ \\
        & B & $20.51$  & 2.180 & $290.8$           \\
\\
V478 Cyg  & A & $16.78$  & 2.185 & $169.8$ & $3.21 $ & $0.223$ &
3.43  & $3.0 \pm 0.3$ \\
        & B & $16.47$  & 2.175 & $154.7$           \\
\\
CW Cep & A & $12.87$ & $ 2.106$ & $52.86$ & $1.61$ & $0.178$ &
1.79    &  $1.640 \pm 0.014$ \\  
	& B & $11.88$ & $2.090$ & $37.23$ \\
\\
QX Car & A & $9.257$ & $ 2.122$ & $16.20$ & $0.171$ & $0.170$ &
0.341   &  $0.340 \pm 0.006$ \\  
	& B & $8.479$ & $2.117$ & $10.96$ \\
\\
U Oph & A & $5.198$ & $ 2.266$ & $2.721$ & $1.85$ & $0.0827$ &
1.93    &  $2.2 \pm 0.3$ \\  
	& B & $4.683$ & $2.256$ & $1.549$ \\
\\
$\zeta$ Phe & A & $3.930$ & $ 2.308$ & $0.9756$ & $0.765$  & $0.0624$ &
0.827   &  $1.03 \pm 0.15$ \\  
	& B & $2.551$ & $2.333$ & $0.1315$ \\
\\
IQ Per & A & $3.521$ & $ 2.278$ & $0.4619$ & $0.363$ & $0.0553$ &
0.418   &  $0.40 \pm 0.03$ \\  
	& B & $1.737$ & $2.416$ & $0.0401$ \\
\\
PV Cas & A & $2.815$ & $ 2.321$ & $0.2647$ & $0.538$ & $0.0572$ &
0.597   &  $0.510 \pm 0.011$ \\  
	& B & $2.756$ & $2.323$ & $0.2705$ \\
\\
VV Pyx & A & $2.101$ & $ 2.488$ & $0.1578$ & $0.0215$ & $0.0661$ &
0.0876  &  $0.0039 \pm 0.0012$ \\  
	& B & $2.099$ & $2.488$ & $0.1572$ \\
\\
EK Cep & A & $2.029$ & $ 2.377$ & $0.05895$ & $0.0153$ & $0.0575$ &
0.0728  &  $0.0030 \pm 0.0009$ \\  
	& B & $1.246$ & $1.867$ & $0.04084$ \\
\\
V1143 Cyg & A & $1.391$ & $ 2.351$ & $0.02735$ & $0.0106$ & $0.0823$ &
0.0929  &  $0.00195 \pm 0.00011$ \\  
	& B & $1.347$ & $2.288$ & $0.02657$ \\
\\
\enddata

\tablenotetext{a}{Radii $R$ in solar units.}
\tablenotetext{b}{Multiply tabular value by $ 10^{-4}$.}

\end{deluxetable}

\clearpage
\objectname{EM Car}
\objectname{V478 Cyg}
\objectname{CW Cep}
\objectname{QX Car}
\objectname{CV Vel}
\objectname{U Oph}

\objectname{zeta Phe}
\objectname{IQ Per}
\objectname{PV Cas}
\objectname{AI Hya}
\objectname{VV Pyx}
\objectname{RS Cha}

\objectname{EK Cep}
\objectname{MY Cyg}
\objectname{PV Pup}
\objectname{DM Vir}
\objectname{V1143 Cyg}
\objectname{UX Men}

\end{document}